\documentclass{aa}
\usepackage{txfonts}
\usepackage{graphicx}
\usepackage{epstopdf}
\usepackage{longtable}
\usepackage{lscape}
\newcommand{\ha}{H$\alpha$}
\newcommand{\hii}{H$_{\rm II}$}
\newcommand{\gal}{G345.45+1.50}
\newcommand{\e}{et al.\ }
\begin{document}

\title{The low-mass pre-main-sequence population of Sco OB1
\thanks{Tables~\ref{table-acis}, \ref{table-xmm} and~\ref{table-mstars}
are only available in
electronic form at the CDS via anonymous ftp to cdsarc.u-strasbg.fr
(130.79.128.5) or via http://cdsarc.u-strasbg.fr/viz-bin/qcat?J/A+A/}
}

\date{Received date / Accepted date}

\author{Francesco Damiani\inst{1}
}
\institute{INAF - Osservatorio Astronomico di Palermo G.S.Vaiana,
Piazza del Parlamento 1, I-90134 Palermo, ITALY
}

\abstract{
The low-mass members of OB associations, expected to be a
major component of their total population, are in most cases poorly
studied because of the difficulty of selecting these faint stars in
crowded sky regions. Our knowledge of many OB associations relies
only on a relatively small number of massive members.
}{
We study here the Sco~OB1 association, with the aim of a better
characterization of its properties such as global size and shape, member
clusters and their morphology, age and formation history, and total mass.
}{
We use deep optical and NIR photometry from the VPHAS+ and VVV surveys,
over a wide area ($2.6^{\circ} \times 2.6^{\circ}$), complemented by
Spitzer IR data, and Chandra and XMM-Newton X-ray data. A new technique
is developed to {find clusters of} pre-main-sequence M-type stars
{ using}
suitable color-color diagrams,
which complements existing selection techniques using narrow-band \ha\
photometry or NIR and UV excesses, and X-ray data.
}{
We find a large population of approximately 4000 candidate low-mass
Sco~OB1 members, net of field-star contaminations, whose spatial
properties correlate well with those of \ha-emission, NIR-excess,
UV-excess, and X-ray detected members, and unresolved X-ray emission.
The low-mass population is spread among several interconnected subgroups:
they coincide with the \hii\ regions \gal\
and IC4628, and the rich clusters NGC~6231 and Trumpler~24, with an
additional subcluster intermediate between these two.
The total mass of Sco~OB1 is estimated to be $\sim 8500 M_{\odot}$.
Indication of
a sequence of star-formation events is found, from South (NGC~6231)
to North (\gal). We suggest that the diluted appearance of Trumpler~24
indicates that the cluster is now dissolving into the field, and that
tidal stripping by NGC~6231 nearby contributes to the process.
}{}


\titlerunning{The PMS population of Sco OB1}
\authorrunning{F.Damiani}

\maketitle

\section{Introduction}
\label{intro}

OB associations are groups of massive stars, often covering
tens of square degrees on the sky. While their massive members have been
intensively studied, low-mass members are generally poorly known. This
is due more to the technical challenges involved in the identification of
these relatively faint stars, than to lack of scientific interest. Most
OB associations are found at low galactic latitudes, projected against
a crowded field-star background; they are distant several hundreds
parsecs, so deep photometry is required to detect their low-mass members.
These latter stars, supposed to be coeval to the massive ones, must
still be in their pre-main-sequence (PMS) evolutionary phase, since O
and early-B stars cannot be older than a few Myr.

Dramatic advances were made in the last two decades with the advent of
sensitive X-ray observations (especially {with} Chandra and XMM-Newton),
since low-mass PMS stars are 3-4 orders of magnitude
brighter in X-rays than older, main-sequence (MS) stars of the same
mass. Therefore, a cluster of PMS stars is very conspicuous in X-ray
images against the field-star population. Several massive clusters, often
found near the central, densest part of OB associations, were studied
in X-rays, and their rich low-mass PMS population identified. Besides
the Orion Nebula Cluster in Ori OB1 (Flaccomio \e 2003a,b, Getman \e 2005),
we mention NGC~2264 in Mon~OB1 (Ram\'{\i}rez \e 2004, Simon and Dahm 2005,
Flaccomio \e 2006),
NGC~6530 in Sgr~OB1 (Damiani \e 2004), NGC~6611 and NGC~6618 (M17) in Ser~OB1
(Linsky \e 2007; Guarcello \e 2007; Broos \e 2007),
NGC~2362 (Damiani \e 2006a), NGC~2244 in Mon OB2 (Rosette Nebula; Wang \e
2008), NGC~6193 in Ara~OB1 (Skinner \e 2005),
Trumpler~14 and~16 in Car~OB1
(Albacete-Colombo \e 2008, Townsley \e 2011), Cyg~OB2 (Albacete-Colombo
\e 2007, Wright \e 2014), NGC~1893 in Aur~OB2 (Caramazza \e 2008),
Cep OB3 (Getman \e 2006),
NGC~6231 in Sco~OB1 (Damiani \e 2016), and several others;
see Damiani (2010) for a review.
In each case,
hundreds {of} low-mass stars were found,
{and up to nearly 10,000 in extremely large regions like Carina}.
{Comparative studies involving several clusters were also made by
Feigelson \e (2013), Getman \e (2017), and Townsley \e (2014, 2018).
}
Studies of the initial-mass-function
on a wide mass range, the total cluster census, star-formation history
and other properties were made possible by these observations.

However, the long exposure times required, and the small field-of-view
(FOV) of these X-ray instruments, compared to the large spatial size
of the OB associations, imply that only the stellar population of their
densest central parts could be studied in X-rays. Sensitive X-ray studies
of low-mass stars across the entire size of an OB association are lacking,
and not even practically feasible with current {X-ray} instruments.

Other valuable means of studying PMS populations spread over wide sky
areas exist: for example, deep narrow-band \ha\ imaging surveys such as
IPHAS (Drew \e 2005) or its southern equivalent VPHAS+ (Drew \e 2014)
have the potential {for} uncovering all stars with strong \ha\ emission
lines, including the classical T~Tauri stars (CTTS), actively accreting
PMS stars. Also the deep near-IR (NIR) surveys of the galactic plane,
such as the UKIDSS (Lawrence \e 2007) or the VVV (Minniti \e 2011) surveys
are providing deep NIR catalogs, from which PMS stars with NIR emission
excesses originated in circumstellar disks (Class~II stars) can be found.
Class~II and CTTS stars (which largely overlap) constitute an important
part of the low-mass content of star-formation regions (SFRs) and young
PMS clusters, but rarely a dominant one. In fact, many PMS stars cease
to accrete matter, and dissipate their disks within a few Myr (Haisch \e
2001), becoming Class~III (or weak-line T~Tauri, WTTS) stars
{ well before the end of their PMS stage}. These latter {stars}
are still X-ray bright, possess active chromosphere and high lithium
abundance, which are straightforward spectroscopic indicators of youth,
but are difficult to find exclusively using optical and NIR photometry.

\begin{figure}
\resizebox{\hsize}{!}{
\includegraphics[]{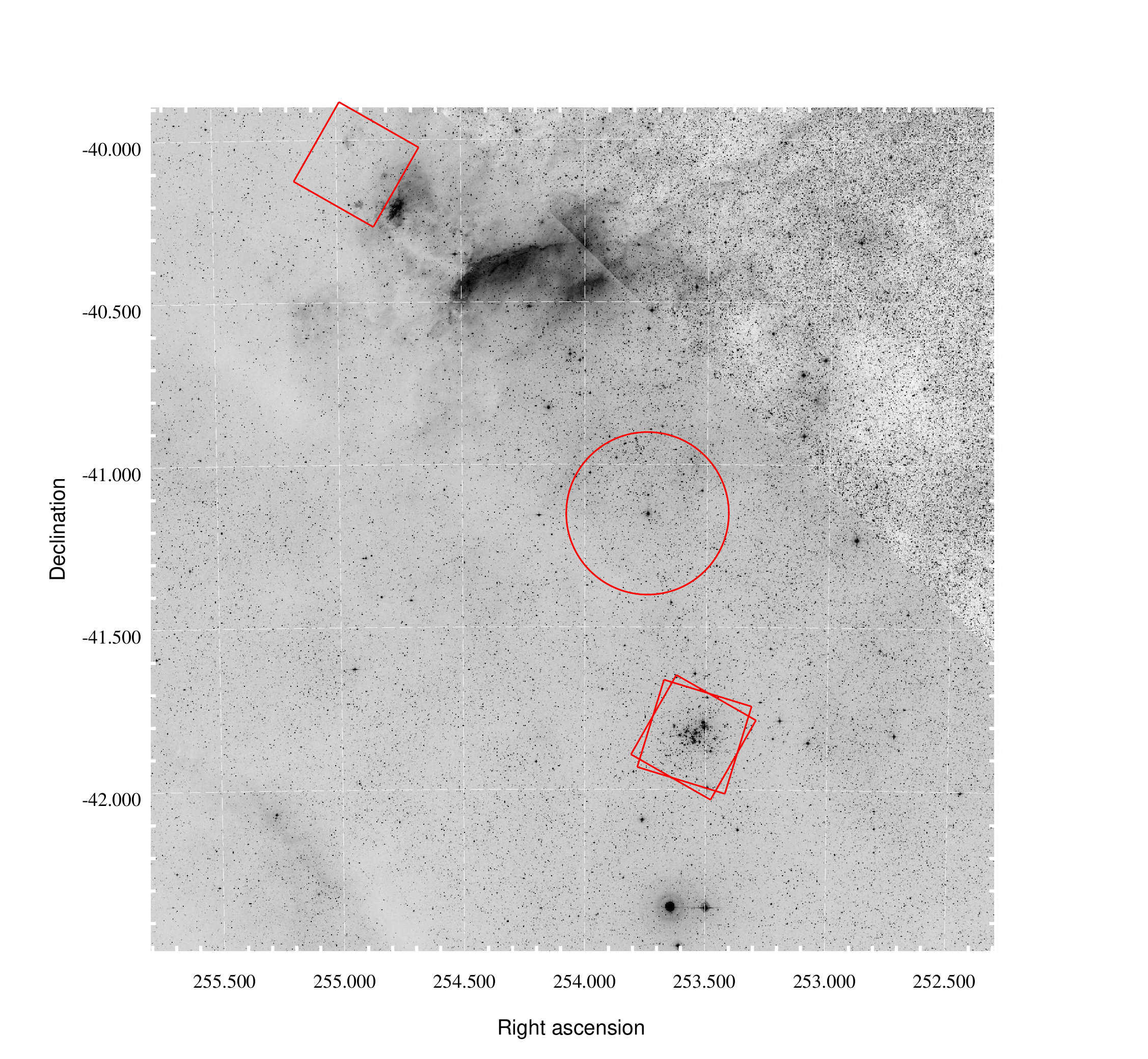}}
\caption{
Digital Sky Survey (DSS2, red band) image of the northern part of Sco~OB1.
The image size is $2.6^{\circ} \times 2.6^{\circ}$ and
coordinates are J2000. The FOVs of existing X-ray images are
indicated with red squares (Chandra) and a circle (XMM-Newton).
The massive cluster NGC~6231 is well visible inside the two southern
Chandra fields. The northern Chandra field is centered on the IR source
IRAS~16562-3959.
Near the middle, the larger XMM-Newton FOV is centered on the Wolf-Rayet
star WR79a, in the Tr~24 region.
The bright nebula near top center is IC4628.
\label{dss-red}}
\end{figure}

Our {little} ability to identify PMS stars outside of dense clusters is
a great obstacle in any study of the long-term evolution of a SFR, and
of its past star-formation history. Lada
and Lada (2003) argue that, on the basis of statistics of embedded and
evolved clusters, the majority of young clusters dissolves already before
10~Myr. This implies that there must be huge numbers of low-mass PMS stars
in the field (maybe 1/1000 of all field stars on the galactic plane). We
are dramatically unable to find them, except that serendipitously in
spectroscopic studies of field stars.  In the early years after the launch
of the X-ray satellite ROSAT several studies were indicating that bright
X-ray sources, found over hundreds square degrees around the Taurus-Auriga
SFR, might be older PMS stars, ejected from Tau-Aur and now dispersing
in the field (Sterzik and Durisen 1995, Feigelson 1996).
These studies were based on
the ROSAT all-sky survey (RASS), which had however too low sensitivity
for {the study of} other, more distant SFRs like the vast majority
of OB associations.
Therefore, there is currently a significant lack of observational data
on both the wide-area low-mass population of OB associations,
and the possible halos of low-mass stars drifting away from SFRs,
which may reveal their past star-formation histories.

In this work we explore an alternative method to those discussed above,
able to find selectively the lowest-mass PMS population in and around a
SFR, of stars with M spectral type. This method exploits the properties of PMS
M stars, and relies on deep, spatially uniform, multiband optical and NIR
photometry, such as that offered by the VPHAS+ (or IPHAS or SDSS, for
example) survey in the optical, complemented by
deep NIR photometry if available.
Despite the faintness of these stars, their
study offers several advantages: they are the most numerous constituent
of a SFR, and provide therefore a member sample with high statistics;
they have a small mass, and are most easily ejected from their parent
cluster by dynamical interactions, thus probing its dynamical evolution.
Finally, they constitute an unbiased sample with respect to presence of
accretion or circumstellar disks.

\begin{figure}
\resizebox{\hsize}{!}{
\includegraphics[]{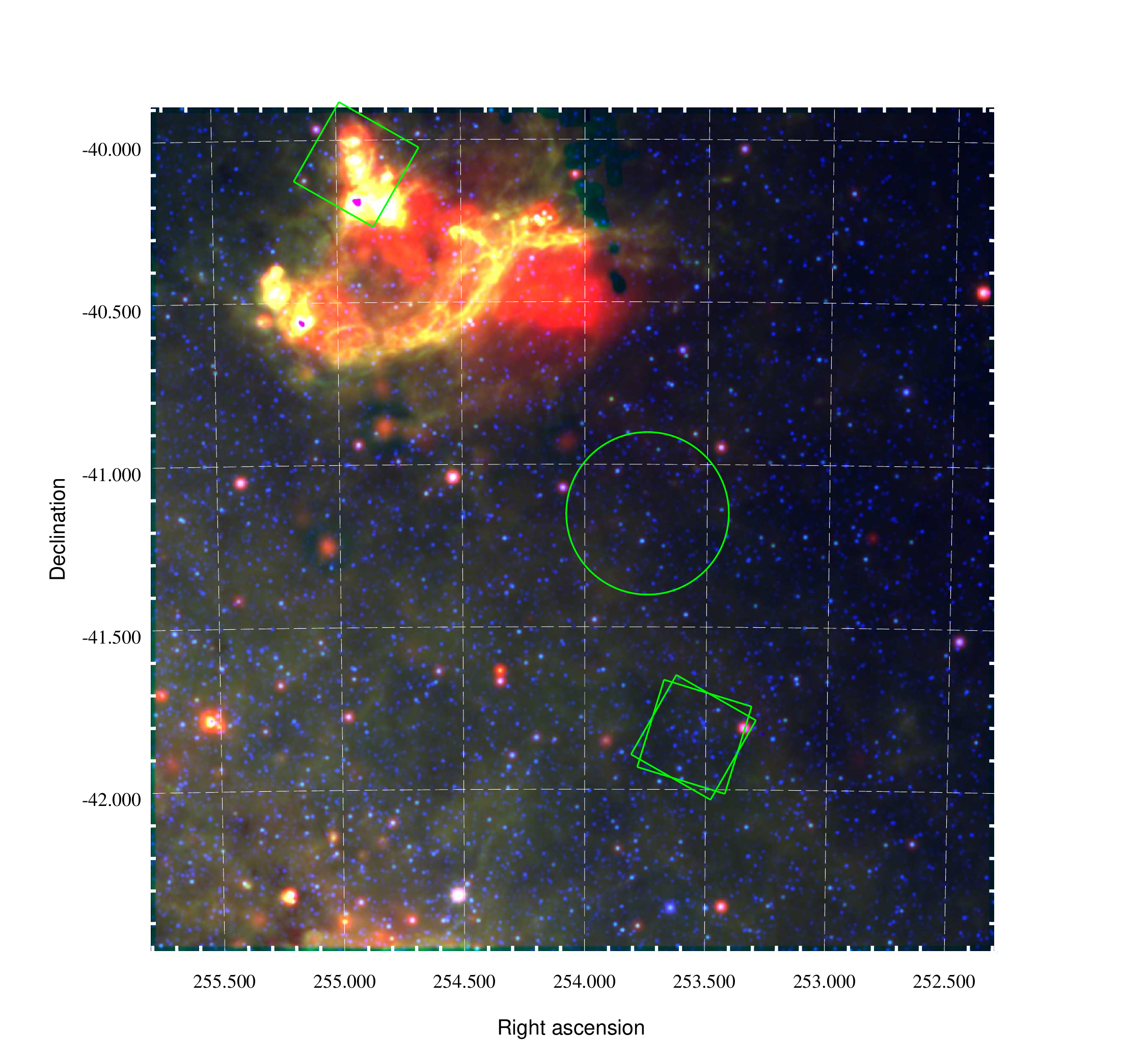}}
\caption{
WISE image of the same sky region as in Fig.~\ref{dss-red}, using bands
at wavelengths $4.6\mu$ (blue), $12\mu$ (green), and $22\mu$ (red).
The same X-ray FOVs as in Fig.~\ref{dss-red} are indicated as a
reference.
The bright diffuse emission coincides with the \hii\ region \gal.
\label{wise}}
\end{figure}

We study here the Sco~OB1 association
(Figures~\ref{dss-red} and~\ref{wise}), using this and other techniques.
The general properties of this large OB association,
which spans almost $5^{\circ}$ on the sky, and is surrounded by a
ring-shaped \hii\ region called Gum~55, are reviewed by Reipurth (2008).
Its central
cluster NGC~6231 contains several tens OB stars, which have been extensively
studied; on the other hand, much fewer studies, all recent, were devoted
to the full mass spectrum, using optical photometry (Sung \e
1998, 2013) and X-rays (Sana \e 2006, 2007, Damiani \e 2016,
{Kuhn \e 2017a,b}).
The currently accepted distance of NGC~6231 is approximately
1580~pc, and its age between 2-8~Myr, with a significant intrinsic
spread (Sung \e 2013, Damiani \e 2016). No ongoing star formation is
known to occur in it, however.
Approximately one degree
North of it, also the loose cluster Trumpler~24 (Tr~24) belongs to the
association. There is little literature on this cluster
(Seggewiss 1968; Heske and Wendker 1984, 1985; Fu \e 2003, 2005)
which unlike NGC~6231 lacks
a well-defined center and covers about one square degree on the sky.
Its age is $<10$~Myr according to Heske and Wendker (1984, 1985), who
find several PMS stars, and its distance is 1570-1630~pc according
to Seggewiss (1968).
Other studies of the entire Sco~OB1 association include
Mac Connell and Perry (1969, \ha-emission stars),
Schild \e (1969 - spectroscopy), Crawford \e (1971 - photometry),
Laval (1972a,b - gas and star kinematics, respectively),
van Genderen \e (1984 - Walraven photometry), and Perry \e (1991 -
photometry).
At the northern extreme of Sco~OB1, the partially obscured \hii\ region
G345.45+1.50 and its less obscured neighbor IC4628
were studied by Laval (1972a), Caswell and Haynes (1987), L{\'o}pez \e (2011),
and L\'opez-Calder\'on \e (2016). They contains massive YSOs (Mottram \e 2007),
maser sources (Avison \e 2016), and the IRAS source 16562-3959 with its
radio jet (Guzm\'an \e 2010), outflow (Guzm\'an \e 2011), and ionized wind
(Guzm\'an \e 2014), and are therefore extremely young (1~Myr or less).
The distance of \gal\ was estimated as 1.9~kpc by Caswell and Haynes (1987),
and 1.7~kpc by L\'opez \e (2011),
in fair agreement with distances of Sco~OB1 stars.
{In Figure~1 of Reipurth (2008) a strip of blue stars is visible,
connecting NGC~6231 to the region of IC4628.}

This paper is structured as follows: Section~\ref{data}
describes the observational data used. Section~\ref{tech}
develops our techniques. Section~\ref{results} presents the results
obtained. Section~\ref{discuss} is a discussion of their significance
in a more general context, and Section~\ref{concl} is a concluding summary.

\section{Observational data}
\label{data}

The spatial region selected for our study is a square of size
$2.6^{\circ} \times 2.6^{\circ}$, centered on $(RA,Dec) = (254.05,-41.2)$
(J2000), that is $(l,b) = (344.19,1.28)$.
This region covers the northern half of Sco~OB1, which comprises the
vast majority of {its}
OB stars, and the bright nebulous region to the North.
The central cluster NGC~6231 is {also} included in our study.

Figure~\ref{wise} shows the same region {as in} Figure~\ref{dss-red},
as seen from the WISE
observatory (Wright \e 2010) in the IR: contrarily to the optical image
of Figure~\ref{dss-red}, NGC~6231 is here barely
noticeable, while a large structure of heated dust is seen to correspond
to the \hii\ region \gal\ and the optical nebula IC~4628.  One of the
brightest part in the WISE image coincides with the source IRAS
16562-3959, targeted by a Chandra observation. The diffuse reddish
emission in the southeast part of the WISE image corresponds to the
Galactic Plane and is probably not related to Sco~OB1, {being
typical of the Galactic equator}.

We primarily use here optical photometric data from the VPHAS+ survey
($ugri$ and \ha\ bands), and NIR photometry from the 2MASS and VVV surveys
($JHK$ bands). These are nearly spatially complete over the region studied.
In addition, we use photometry from the Spitzer Glimpse survey
(Spitzer Science Center 2009),
and X-ray data from the archives of Chandra and XMM-Newton observatories,
which as seen from Figure~\ref{dss-red} cover only a very small fraction of
the studied region.

\subsection{Optical and NIR data}
\label{optnir}

\begin{figure}
\resizebox{\hsize}{!}{
\includegraphics[height=12.0cm]{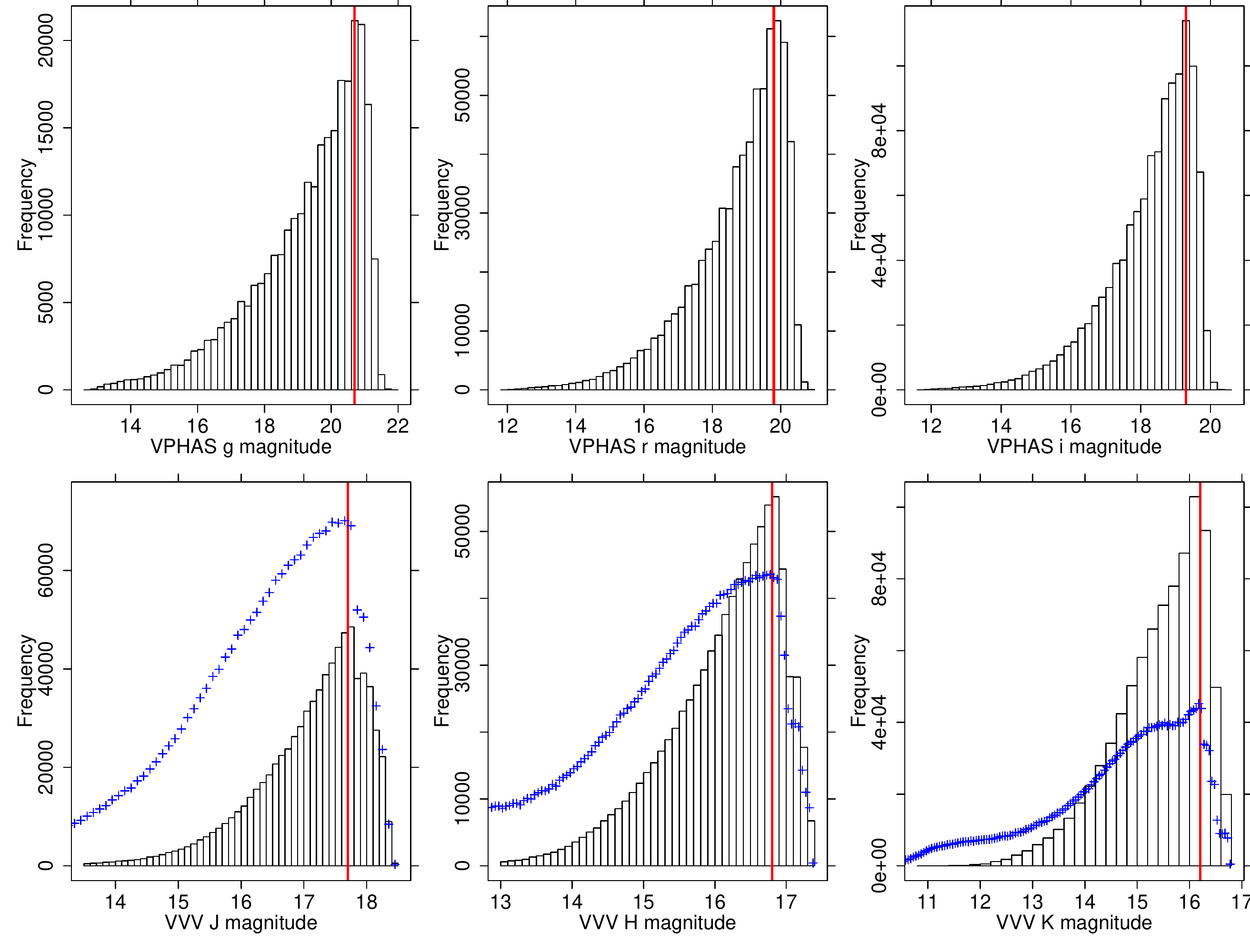}}
\caption{
Magnitude histograms for VPHAS+ bands $g,r,i$ (top) and VVV bands $J,H,K$
(bottom). The vertical red lines indicate our adopted completeness limits.
{ In the lower panels, the magnitude distributions of VVV catalog entries
discarded because of non-zero {\em ppErrBits} flags are indicated by
blue symbols.}
\label{grijhk-hist}}
\end{figure}

In the studied region, the VPHAS+ DR2 data comprise 2014322 individual
sources with at least one of flags {\em clean\_u, clean\_g, clean\_r,
clean\_i, clean\_\ha} equal to 1 (i.e., {a good and reliable magnitude
in the given band}).
Individual $ugri$ and \ha\ magnitudes
were discarded when the corresponding {\em clean} flag is zero, but the
source itself is not discarded {if it has} a good measurement in
another of the VPHAS+ bands.
The VPHAS+ data miss stars brighter than $i \sim 12$, which in Sco~OB1
correspond to massive stars {(above 4 $M_{\odot}$)},
thoroughly studied in the literature;
since the emphasis here is on low-mass Sco~OB1 members, this limitation has
no consequences on our study.
Histograms of the VPHAS+ source magnitudes in the $gri$ bands are shown
in Figure~\ref{grijhk-hist} (top panels). From these we determine
completeness limits of $g=20.7$, $r=19.8$, and $i=19.3$, respectively.
In regions with nebulosity the local limiting magnitude may be
slightly brighter, depending on the nebular intensity in the given band,
because of the increased background level.
This might be of some consequence only in the vicinity of IC~4628
(Fig.~\ref{dss-red}), but for most of the studied region the VPHAS+ limiting
magnitudes are spatially uniform.
The spatial completeness of these optical data across the region is
examined in detail in Section~\ref{results}.

We used NIR data from VVV DR4 (table {\em vvvSource}), selecting only
sources with stellar or probably stellar characteristics (flag {\em
mergedClass} equal to $-1$ or $-2$). These comprise 8142990 sources, of
which only the $J$, $H$, and $K_s$ band magnitudes are considered here
({ we disregarded} VVV $Y$ and $Z$ bands, {as they were not useful
for this study}).
However, tests on NIR color-magnitude and color-color diagrams
showed thousands of sources falling in unexpected places,
{ and most likely having spurious colors}.
These were as a rule found to possess flags {\em jppErrBits}~$>0$, {\em
hppErrBits}~$>0$, or {\em ksppErrBits}~$>0$. We therefore disregarded
all magnitudes with the corresponding flag set to a nonzero value. This
is a stronger filtering than suggested in the VVV DR4 explanatory notes
(i.e., disregard only sources with {\em ppErrBits} flags greater than
255), but we checked that setting the flag threshold to 255 (or even
down to 16) would result in too many sources with peculiar (and likely
spurious) NIR colors, and therefore false positives in our selection
(Section~\ref{halpha}). Moreover, filtering by magnitude errors did not
improve the selection, since many suspicious flagged sources are
characterized by very small formal errors.
After removing VVV sources with all three flags
{\em jppErrBits}~$>0$, {\em hppErrBits}~$>0$, and {\em ksppErrBits}~$>0$
we obtain a final number of 2663602 clean VVV sources, that is
approximately one-third of the initial catalog.
{ Such a strong filtering likely excludes many thousands sources with
only minor photometric inaccuracies (along with those with major
accuracy problems)}
and therefore this step introduces a significant incompleteness in
the VVV source catalog (with little magnitude dependence),
probably by more than a factor of two.
{ However, we checked that this does not introduce a
spatial-dependent bias, which would be most harmful in our context.}
The limiting magnitudes for the VVV bands are derived from the
histograms in Figure~\ref{grijhk-hist} (bottom panels), as $J=17.7$,
$H=16.8$, and $K_s=16.2$, respectively, for sources with magnitude errors
less than 0.1~mag.
{ The same panels also show the magnitude distributions of discarded
VVV sources: the discarded/good source number ratio increases towards the
bright magnitudes, as is typical of systematic errors and the opposite
to problems caused by low S/N; also the detailed shapes of their distributions,
especially that in the $K$ band, suggest that these magnitudes are not
very reliable and are better ignored.}

Figure~\ref{grijhk-hist} also shows the bright
limits for the VVV catalog, at $J \sim 14$, $H \sim 13$ and $K_s \sim 12$.
Many Sco~OB1 members, not only of OB types, are expected to be brighter
than these limits.
Therefore,
we need to complete the VVV NIR dataset above its bright limit, for a
proper study of Sco~OB1 low-mass stars. The most
natural choice for this magnitude range is the 2MASS catalog (Cutri \e
2003), complete approximately down to $J \sim 15$, $H \sim 14$ and $K_s
\sim 13.8$, {once} we require magnitude errors less than 0.1~mag.
We disregarded individual 2MASS magnitudes in the presence of at least one of:
flag {\em ph\_qual} equal to 'E', 'F', 'U', or 'X';
flag {\em rd\_flg} outside the range [1-3];
nonzero {\em cc\_flg} flag;
flag {\em bl\_flg} different from 1.
This choice is more conservative than that adopted in some previous works
(e.g., Damiani \e 2006b, where we kept sources with confusion flag
{\em cc\_flg='c'}), and undoubtedly misses some real sources in dense
regions\footnote{{ In practice, this has some consequences for this
study only in the inner parts of NGC~6231.}}.
However, since our present approach relies on accurate optical and NIR
colors, we opted for the most conservative filtering, both on 2MASS and
VVV data.
There are 717232 {clean} 2MASS sources in the studied region.

We assembled a combined NIR catalog
by positionally matching the VVV and 2MASS catalogs using a maximum
distance of 0.2~arcsec. This yielded 118612 sources common to both 2MASS and
VVV, and 3262222 unique NIR sources {(the intersection and union of
the two catalogs, respectively)}.
The common magnitude range between 2MASS and VVV in each of the $J$, $H$,
and $K$ bands enabled us to calibrate the VVV photometry in the 2MASS system
(see Appendix~\ref{append1}). For sources detected in both datasets we
selected the 2MASS magnitudes, to avoid an excessive number of missing
values because of {\em ppErrBits} flagged entries in the VVV dataset.
We examine in detail the spatial completeness of the
combined NIR catalog in Section~\ref{results}.

\begin{figure}
\resizebox{\hsize}{!}{
\includegraphics[]{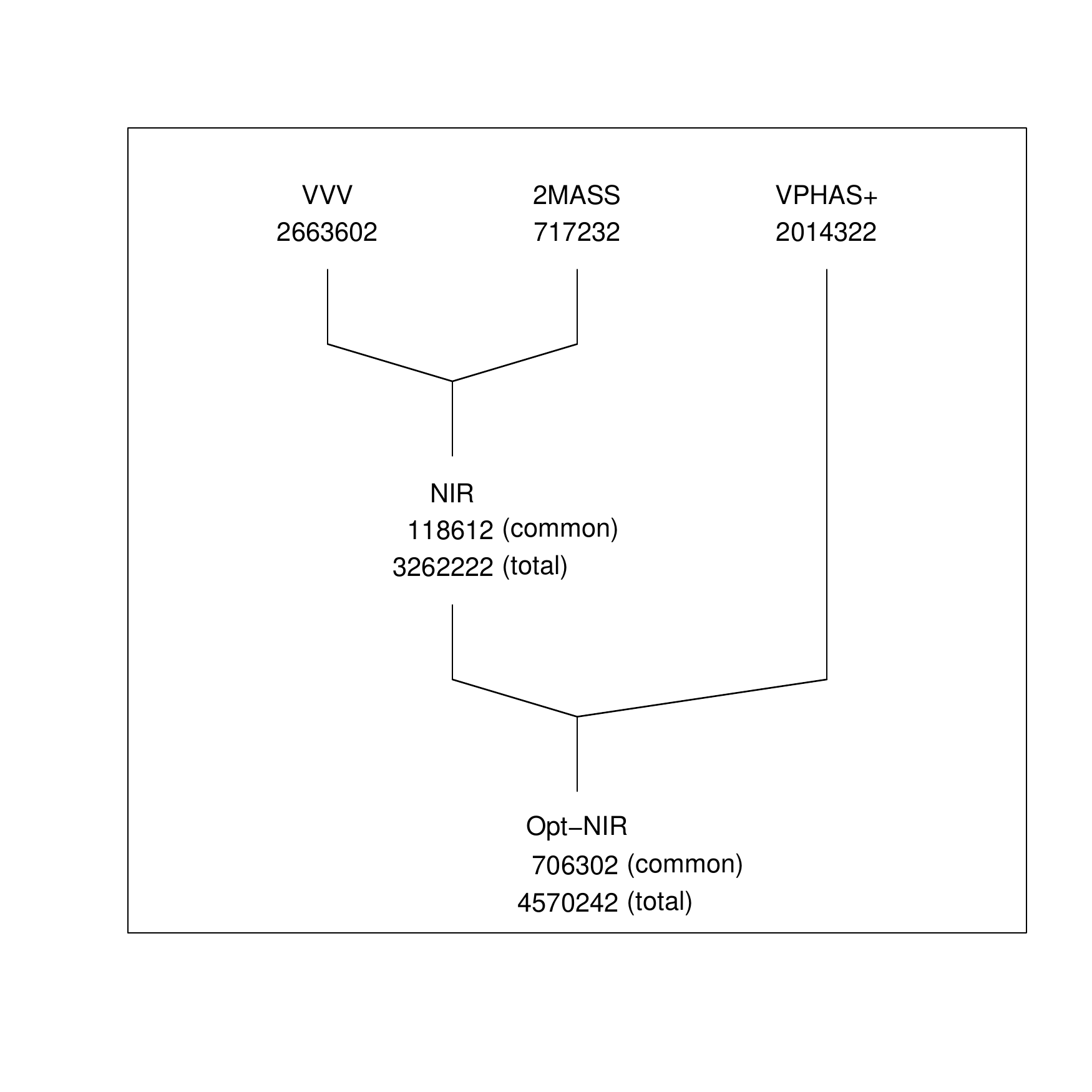}}
\caption{
 Number of stars in each input catalogs after cleaning, and
construction of NIR and Optical-NIR catalogs after source matching.
Labels 'common' and 'total' refer to the intersection and union of input
catalogs, respectively.
\label{mastercat}}
\end{figure}

Using this combined NIR catalog and the VPHAS+ catalog we finally produced an
optical-NIR catalog, by positionally matching sources within 0.2~arcsec.
The {total} number of entries in this optical-NIR catalog is 4570242
{ (the
union of input catalogs, including unmatched entries)},
{ while the number of stars detected in both the optical and the NIR
is 706302 (intersection of input catalogs); see Figure~\ref{mastercat}}.

The Glimpse Source catalog (I + II + 3D, Spitzer Science Center 2009,
available from CDS, Strasbourg)
contains 719188 IR sources in our spatial
region. Of these, we consider the subset (52740 sources ) having magnitude
errors less than 0.07~mag in each of the [3.6], [4.5], [5.8], and [8.0] bands.

\subsection{X-ray data analysis}
\label{pwdet}

Besides the deep Chandra X-ray data on NGC~6231 presented in Damiani
\e (2016), we used the Chandra archive data on IRAS~16562-3959.
This observation, made with the ACIS-I detector, was
split in three pointings (ObsIDs 14537, 17691, and 16658; exposure
times 5017~s, 40066~s, and 38576~s, respectively; PI: P.\ Hofner),
nearly co-pointed and with total exposure time 83659~s.
We have detected sources in the three pointings (co-added) using our
detection code PWDetect (Damiani \e 1997a,b), which is capable of
dealing with combined Chandra datasets (see, e.g., our study of NGC~2516,
Damiani \e 2003). The chosen detection threshold corresponds
approximately to one spurious detection in the FOV. We found 384 point
sources in the IRAS~16562-3959 combined dataset, as listed in
Table~\ref{table-acis}. The spatial distribution of these sources and
their connection with stars detected at other wavelengths are discussed
in Section~\ref{results}.

\begin{figure*}
\resizebox{\hsize}{!}{
\includegraphics[]{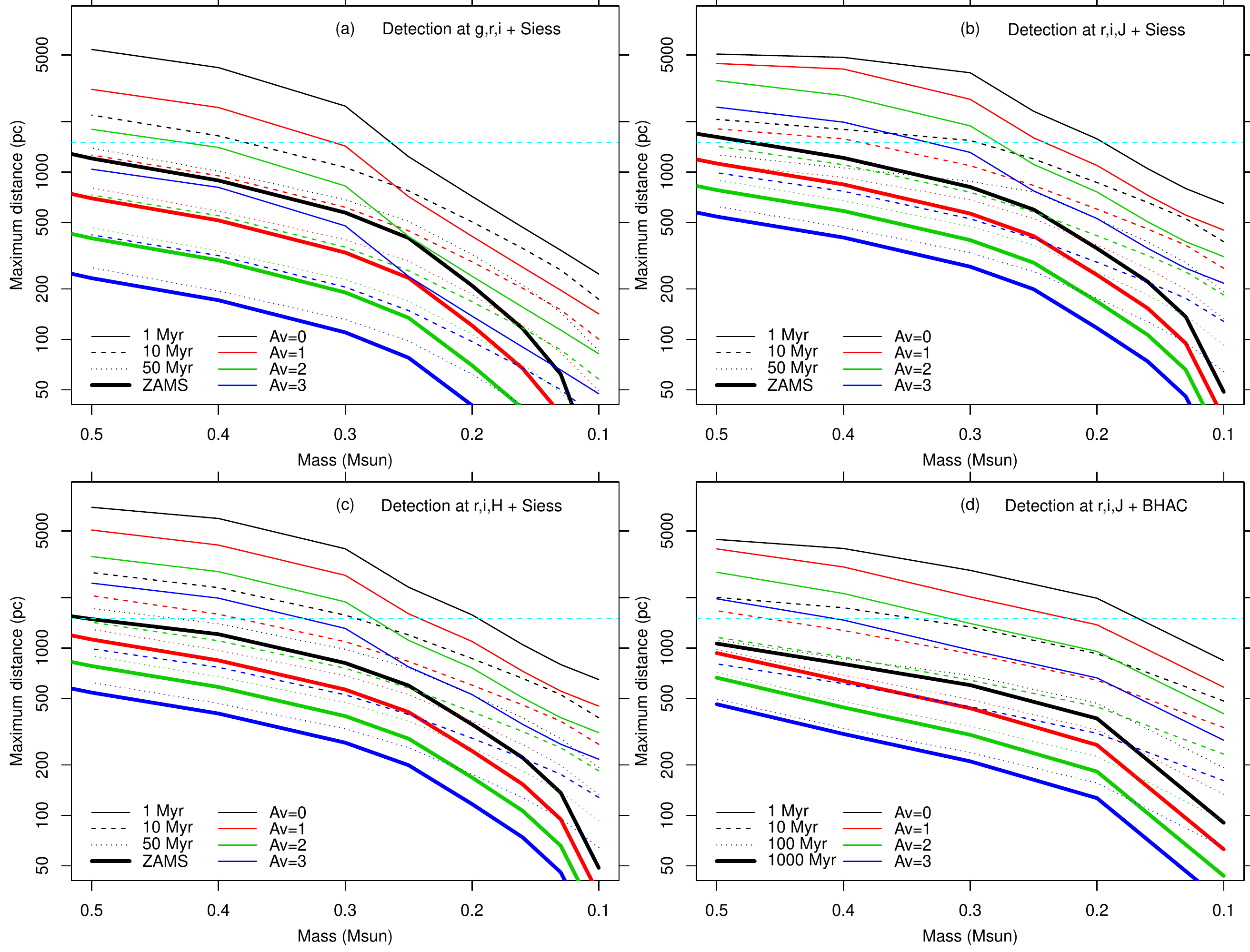}}
\caption{
{ Mass-Distance-Age (MDA) diagrams:}
Maximum distances $d_{XYZ}^{max}$ for simultaneous detection in three bands,
using completeness limits from Fig.~\ref{grijhk-hist}, and PMS evolutionary
tracks from Siess \e (2000) or BHAC, as indicated. The two right panels
permit evaluation of the model-dependent uncertainties in the maximum
distances. The cyan horizontal dashed lines indicate the distance of
the NGC~6231 cluster.
\label{max-distance}}
\end{figure*}

Relevant X-ray data on Sco~OB1 are also found in the XMM-Newton archive, in
addition to those on NGC~6231 published by Sana \e (2006). We consider
the XMM-Newton observation of the Wolf-Rayet star WR~79a
(ObsID 0602020101, exposure 36080~s, PI: S.\ Skinner), which also covers
part of Tr~24. We used our code PWXDetect (that is, the version
optimized for XMM-Newton data) to detect sources in the combined images from the
three detectors in the EPIC camera ({\em pn}, MOS1, and MOS2). This code
version was already used in previous cluster studies such as for example
Sciortino \e (2001). We detected in this way 195 X-ray sources, listed
in Table~\ref{table-xmm}, and discussed in Section~\ref{results}.

\section{Techniques for selection of young stars}
\label{tech}

{
In this section we discuss the various methods we use to select
candidate PMS stars. We first present our method to photometrically
select M stars, and the reasons why this is especially useful when
applied to PMS stars. Then, other widely used photometric methods, such
as those based on \ha, IR and UV excess emission, are also used to establish
reference PMS star samples for the same sky region, which help us to
test the effectiveness and reliability of our selection of M-type PMS stars.
}

\subsection{M stars}
\label{mstars}

Our method for identification of M stars in a SFR is based on a few
key properties of these stars. First, it is commonly accepted that during PMS
evolution stars contract toward the MS; despite significantly different
details between various sets of model evolutionary tracks (such as
Baraffe \e 2015 - BHAC, or Siess \e 2000), there is consensus on the
fact that, {among all spectral types, M-type} stars {(with masses
in the range 0.1-0.5$M_{\odot}$)}
show the largest luminosity excursion between an
age of 1~Myr and their arrival on the ZAMS. BHAC predict a difference
of 3.3 magnitudes for a 0.5$M_{\odot}$ star (i.e., a factor of 21 in
$L_{bol}$), and a difference of 5.0 magnitudes for a 0.1$M_{\odot}$
star (a factor of 100 in $L_{bol}$).  For comparison, a 1$M_{\odot}$ star
evolves by only 1.36 magnitudes (a factor of 3.5 in $L_{bol}$).
Therefore, {for a given limiting magnitude}
a 1~Myr old, 0.5$M_{\odot}$ (0.1$M_{\odot}$) star will be detectable
until a distance 4.6 (10) times larger than a ZAMS star of the same
mass. If a cluster of M-type PMS stars is present in a given sky region,
it becomes {prominent} with respect to the field M dwarfs, detected
only across a much smaller volume.

{
The second relevant property of M stars is that their spectral energy
distributions are shaped in such a way that temperature and reddening
become non-degenerate using suitable pairs of optical and NIR colors,
as we discuss in detail below.}
The temperature-reddening degeneracy is instead rather
ubiquitous for all hotter stars, and is {ultimately} responsible for a
large fraction of non-member contamination in photometrically selected
samples of SFR members.

We examined quantitatively the applicability of this selection method,
for the VPHAS+ and VVV magnitude limits (Fig.~\ref{grijhk-hist}).
We considered isochrones from Siess \e (2000) at different ages up
to the ZAMS, converted to the VPHAS+ photometric system according to the
prescriptions given in Drew \e (2014).
These isochrones, together with the above magnitude limits, were used to
predict the maximum distance $d_X^{max}$ for detection of a star of given
mass and age in each individual photometric band, say $X$.  The maximum
distance $d_{XYZ}^{max}$ for simultaneous detection in a given set of
bands $XYZ$, for a star
to be placed in the $(X-Y,Y-Z)$ color-color diagram, is equal to the minimum
value among the
limiting distances for the respective bands:
$d_{XYZ}^{max}= min (d_X^{max}, d_Y^{max}, d_Z^{max})$.
We also computed $d_{XYZ}^{max}$ for non-zero extinction, of $A_V = 0,1,2,3$
mag., which gave (obviously) gradually decreasing distances $d_{XYZ}^{max}$
toward the higher extinction values.  These results are shown
graphically in Figure~\ref{max-distance} for the band combinations
$(g,r,i)$, $(r,i,J)$, and $(r,i,H)$. The reason behind these choices of
bands is discussed below. A fourth plot uses again the $(r,i,J)$ bands,
but this time together with the BHAC isochrone set, to permit evaluation of
the importance of model-related uncertainties
{(a comparison involving the $(r,i,H)$ bands gives a similar result)}.
The distance of NGC~6231 is shown as a reference.
{ These diagrams, henceforth referred to as Mass-Distance-Age (MDA)
diagrams, are of greatest importance for all the following
development, and deserve a detailed analysis.}

\begin{figure*}
\includegraphics[width=16.6cm]{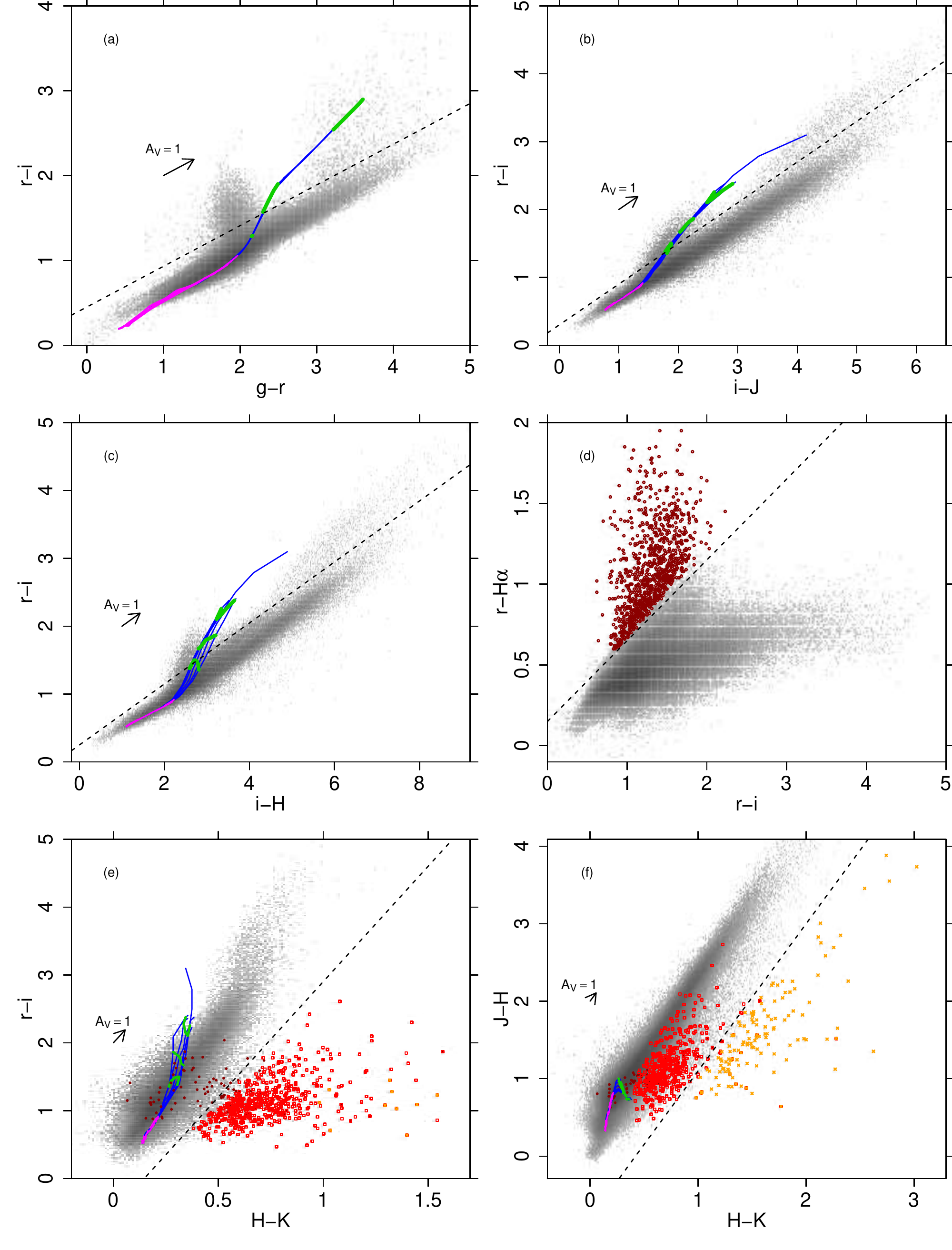}
\caption{
Color-color diagrams used for selection of young stars.
Each diagram is a 2-dimensional histogram, with shades of gray indicating the
density of datapoints.
Only data with errors less than 0.1~mag on each pair of colors are shown.
Panels $(a)$,
$(b)$ and $(c)$: M stars are found above the dashed lines. Panel $(d)$:
strong \ha-emission stars are found above the dashed line (dark-red points).
Panels $(e)$ and $(f)$: IR-excess stars are found below the dashed
lines (red squares: selected from $(H-K,r-i)$ diagram; orange crosses:
selected from $(H-K,J-H)$ diagram; dark-red points from panel $(d)$).
Representative reddening vectors are shown, except in panel $(d)$
where reddening moves datapoints along curved trajectories. BHAC
isochrones (evolutionary tracks) are shown with blue (green) lines for
ages 1, 10, 50 and 10,000~Myr (masses of 0.1, 0.3, and 0.5 $M_{\odot}$),
reddened as appropriate for NGC~6231 {($A_V = 1.5$)}.
The location of stars with mass $M>1 M_{\odot}$ (all ages) is indicated with a
magenta curve. In panel $(a)$ only, Siess \e
(2000) isochrones and tracks are used, the BHAC set being unavailable for the
$g$ band.
\label{definitions}}
\end{figure*}

\begin{figure}
\resizebox{\hsize}{!}{
\includegraphics[]{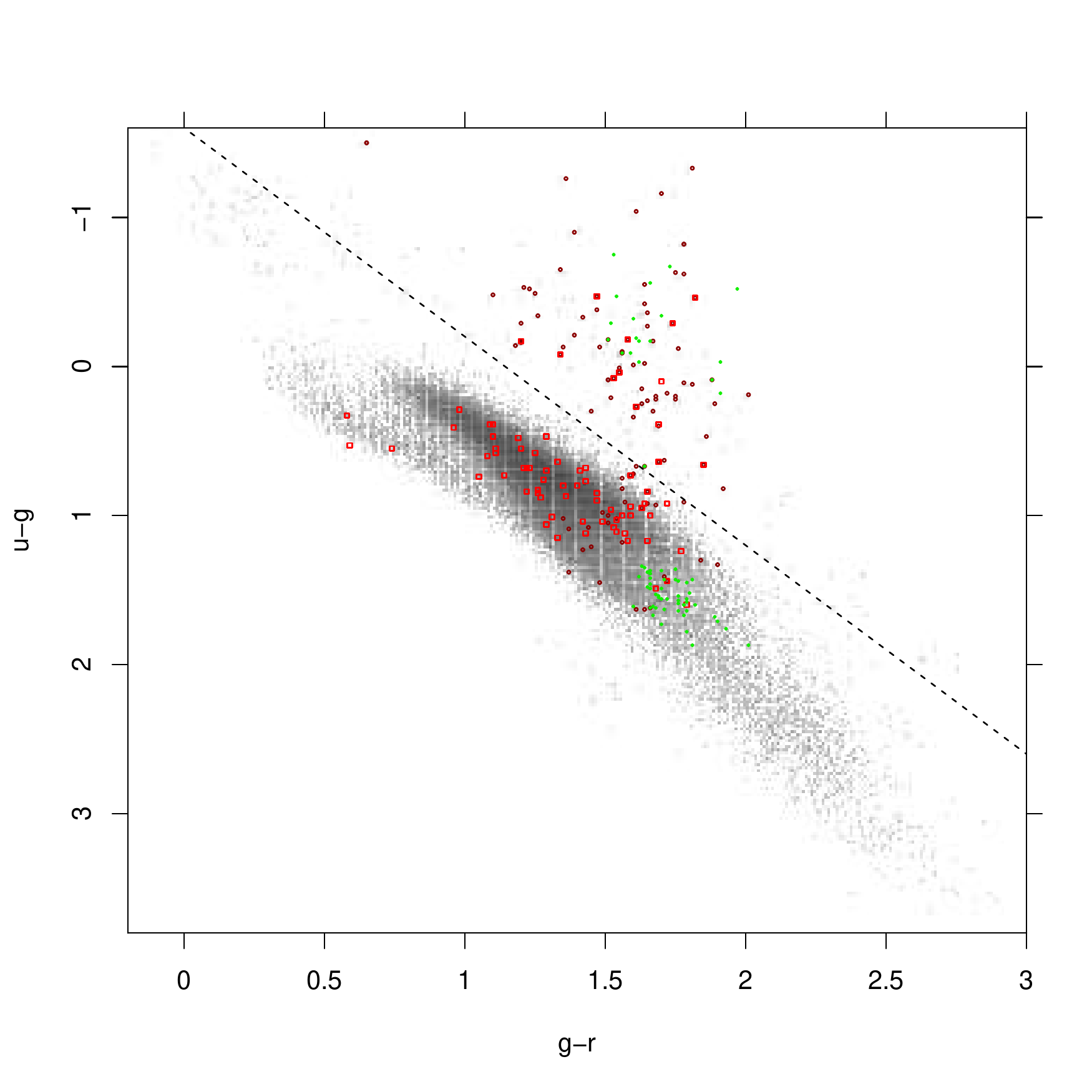}}
\caption{
A $(g-r,u-g)$ color-color diagram used for selection of UV-excess stars.
Colored symbols as in Figure~\ref{definitions}, with the addition of
green dots indicating M stars. UV-excess stars are found above the dashed
line.
\label{ug-gr}}
\end{figure}

{ The $(g,r,i)$ MDA diagram of}
Figure~\ref{max-distance}$a$ shows that with the available VPHAS+
data we are able to detect M stars, at the distance of NGC~6231 and
extinction $A_V=1$ mag, down to a mass $\sim 0.3 M_{\odot}$,
simultaneously in $g$, $r$, and $i$. Down to the same mass limit, MS M
dwarfs are detected only up to $\sim 600$~pc, or a space volume 15 times
smaller. The use of the $g$ band might be a limiting factor for PMS
stars, if they are found in obscured regions or dusty local environments;
therefore, we examined the {$(r,i,J)$ MDA diagram,
also shown in Figure~\ref{max-distance}$b$.}
Here the distance between curves at different $A_V$ values is reduced,
as expected since the bands used are redder and less affected by extinction.
For the NGC~6231 distance and $A_V=1$ we detect stars down
to {a minimum mass of} $\sim 0.25 M_{\odot}$, which rises only slightly to $\sim 0.28
M_{\odot}$ for $A_V=2$. The limits obtained (Figure~\ref{max-distance}$c$)
using the $H$ band
in place of $J$ are almost the same. Finally, using BHAC instead
of Siess \e (2000) models changes only small details in these
predictions, as can be seen from comparing the {two MDA diagrams in the} right panels.
At 10~Myr, the intrinsic luminosity of PMS M stars diminishes, and with
it also the contrast with the field M dwarfs; however, also the average
extinction is often lower at this age with respect to 1~Myr, so the two
effects partially compensate: a 10~Myr, $A_V=0$ star at the NGC~6231
distance is detected in $gri$ down to a mass $\sim 0.4 M_{\odot}$, and
in $riJ$ down to $\sim 0.3 M_{\odot}$.

{ The mere detection of M stars ensured by the above discussion,
against a very crowded background of stars would be of little use, if
not accompanied by a recipe to discriminate M stars from all other, hotter
stars. This is possible since temperature and reddening are here
non-degenerate, as mentioned before, and shown in detail below.}

Our ability to select stars {of M type} using the available
photometric data is demonstrated from the color-color diagrams in
Figure~\ref{definitions}$a,b,c$,
{ where both theoretical isochrones and observational data from our Sco
OB1 optical-NIR catalog are shown.}
Each of these diagrams uses a
combination of three photometric bands, corresponding to {one of the
MDA diagrams}
in Figure~\ref{max-distance}.
The large majority of stars follow a linear locus, parallel to the
reddening vector. Above that, a small number of stars deviate from that
locus in each diagram.  The isochrones and evolutionary tracks shown in
the figure indicate that these deviant stars have colors corresponding
to little-reddened M stars with masses $\leq 0.5 M_{\odot}$,
irrespective of their age between 1~Myr and ZAMS age. More massive stars
show instead degenerate temperatures and reddening in these diagrams.
We have therefore set fiducial limits for discrimination between the M
stars and hotter stars, illustrated by the dashed lines in the figure.
This was done is a rather conservative way, to minimize the number of
false positives, or spuriously selected hotter stars.
Only stars with errors less than 0.1~mag on each color are shown in the
respective diagrams.
In panel $a$ of Figure~\ref{definitions} the shape of the M-star locus
appears different (more vertical) from that of the model locus.  This,
however, may easily be related to a selection effect: The more massive M
stars ($\sim 0.5 M_{\odot}$) are also relatively brighter, and detectable
in the $g$ band also in the presence of moderate reddening.  Instead the
less massive stars ($\sim 0.2 M_{\odot}$) near the top of the group are
much fainter (and probably more spatially dense), and detectable in $g$
only at low reddening.
{In general, comparison between panels $a$ and $b$ shows that M
stars detected in the $g$ band are in general less reddened than those
detected in the $J$ band, as expected.}

At colors $g-r \geq 3$, another loose group of stars can be seen in
the $(g-r,r-i)$ diagram of
Figure~\ref{definitions}$a$ above the M-star limit: these are stars
sharing their intrinsic colors with M dwarfs, however at much larger
reddening. It is natural to identify them with distant M giants, which
are so rare in terms of their absolute volume density {that they are} found
predominantly at large distances (and reddening). This assumption
is confirmed by their {high} observed brightness (median $i=15.9$,
compared to $i=18.0$ for the M dwarfs), despite their larger reddening.
These groups of M giants are also clearly evident in the $(i-J,r-i)$ and
$(i-H,r-i)$ diagrams of Figure~\ref{definitions}$b,c$, and comprise the
highest-reddening stars found in the latter diagrams.
The importance of studying the background M-giant population in relation
to the low-mass PMS population in Sco~OB1 will become clearer in
Section~\ref{results} below.

{ Therefore,} we define as $M_{gri}$ stars those selected using:
\begin{equation}
(r-i) > 0.45+0.48\; (g-r)
\label{mgri}
\end{equation}
{ (that is, above the dashed line in Figure~\ref{definitions}$a$)}
and also $1.5<(g-r)<2.5$ to
discriminate from high-reddening M giants.
Similarly, $M_{riJ}$ stars are defined as:
\begin{equation}
(r-i) > 0.3+0.6\; (i-J)
\label{mrij}
\end{equation}
and $1.4<(i-J)<3.0$.
Also, $M_{riH}$ stars are defined as:
\begin{equation}
(r-i) > 0.25+0.45\; (i-H)
\label{mrih}
\end{equation}
and $2<(i-H)<3.5$.
The M giants are defined as: $gM_{gri}$ from Eq.~\ref{mgri} and $(g-r)>2.5$,
$gM_{riJ}$ from Eq.~\ref{mrij} and $(i-J)>3.5$,
and $gM_{riH}$ from Eq.~\ref{mrih} and $(i-H)>3.5$.
The numbers of M-type stars selected, by type, are 5367 $M_{gri}$, 5133
$M_{riJ}$, and 4190 $M_{riH}$ stars, with considerable overlap
{ among the three groups,} for a
total number of 10224 low-reddening M stars
{(reported in Table~\ref{table-mstars})}. The corresponding numbers
for reddened M giants are 647 $gM_{gri}$, 977 $gM_{riJ}$, and 1447 $gM_{riH}$
stars (total number 1936 stars).
{The overlap between the $gM_{riJ}$ and $gM_{riH}$ samples is nearly
complete, while only 167 stars are common to the $gM_{gri}$ and
$gM_{riH}$ samples.}

{If strong local extinction is present (circumstellar rather than
interstellar dust), typically near the youngest stars, its effect might
be to move an M star into the ``reddened giants'' subsample. We discuss
in Section~\ref{spatial} that this latter subsample does not show spatial
clustering, which argues against local extinction to be relevant for a
significant number of Sco~OB1 M stars.  }

To summarize, we have defined regions in several color-color diagrams
which permit to select effectively samples of M stars;
the PMS M-type members of
a SFR in the 1-10~Myr age range, and over a wide interval of
distances and extinction, are expected to outnumber the
much closer field M~dwarfs {in a magnitude-limited sample}.
The {M stars selected in this way} constitute therefore a good sample
of candidate members of the SFR.

\subsection{\ha\ emission, IR and UV excesses}
\label{halpha}

We considered also other diagnostics of PMS status in this study of
Sco~OB1. The first is photometric \ha\ excess, which is one of the
fundamental products of the VPHAS+ survey. CTTSs are efficiently
selected by using star positions in the $(r-i,r-H{\alpha})$ diagram, as
is done for example in Kalari \e (2015) for the young cluster NGC~6530,
or in our study of the NGC~7000 nebula (Damiani \e 2017).
We use a similar fiducial limit for the present VPHAS+ dataset in
Sco~OB1, shown in Figure~\ref{definitions}$d$:
\begin{equation}
(r-H{\alpha}) > 0.15+0.5\; (r-i)
\end{equation}
Only stars 2$\sigma$ above the fiducial limits were selected.
This yielded 1082 \ha-excess stars in the entire region studied.

\begin{figure}
\resizebox{\hsize}{!}{
\includegraphics[]{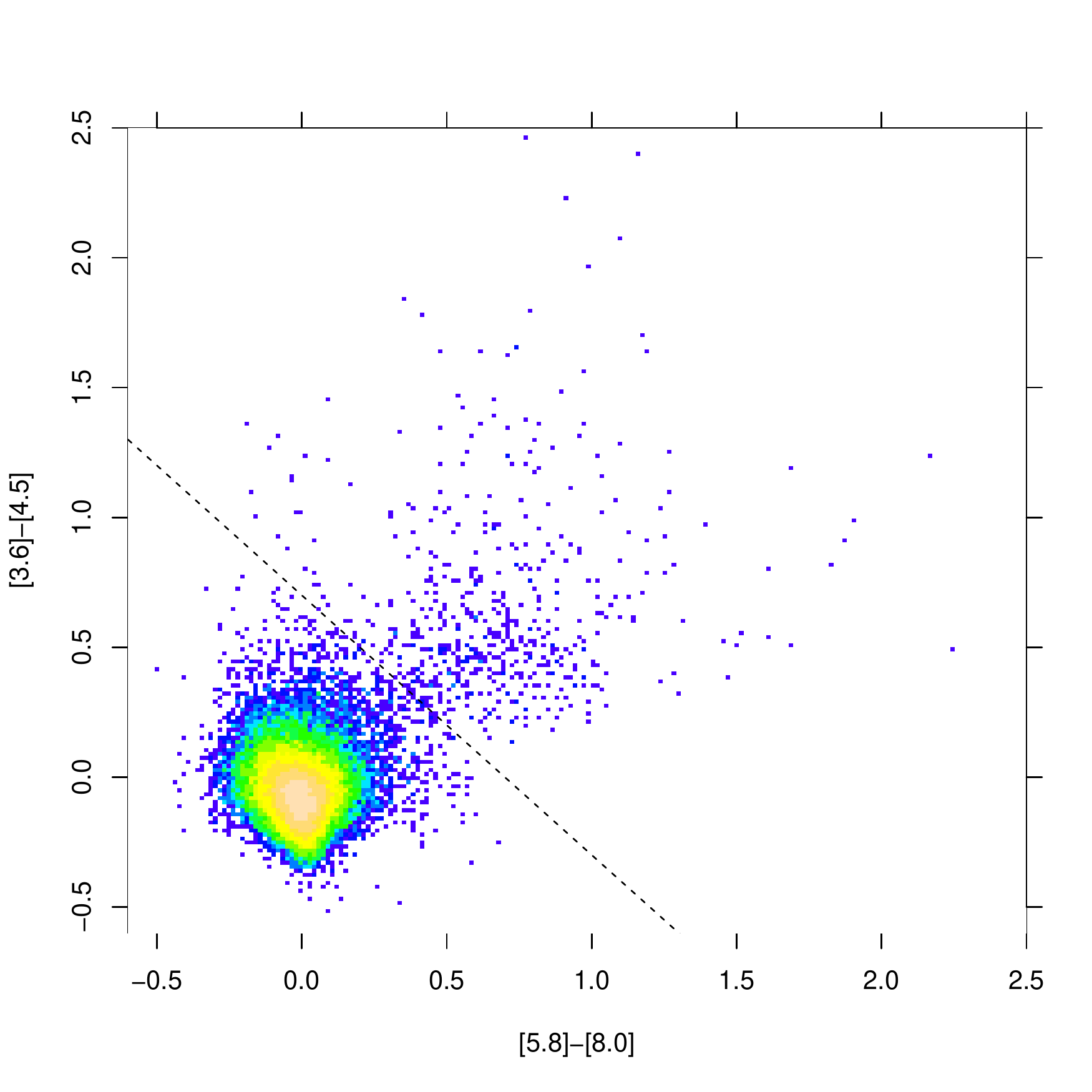}}
\caption{
Spitzer $([5.8]-[8.0],[3.6]-[4.5])$ color-color diagram. Stars above the
dashed line are YSO candidates.
\label{glimpse-col-col}}
\end{figure}

Class~II members of a SFR are efficiently selected thanks to their NIR
excess emission, best detected using the $K$ band. {It was} argued in
previous works (Damiani \e 2006b, Guarcello \e 2007, 2009, Damiani \e 2017;
see also Strom \e 1993, Eiroa and Casali 1995) that a mixed
optical-NIR color-color diagram provides a more effective tool to select
this type of PMS stars than an exclusively NIR-based diagram such as
$(H-K,J-H)$. Therefore, we consider here the $(H-K,r-i)$ diagram, shown
in Figure~\ref{definitions}$e$. As above, only stars with color errors
less than 0.1~mag are shown. We set a fiducial limit, parallel to the
reddening vector, and selected all stars with redder $(H-K)$ colors (above
$2\sigma$ significance) with respect to that limit:
\begin{equation}
(r-i) < -0.5+3.4\; (H-K)
\end{equation}
which yielded 849 optical-NIR excess stars.

\begin{figure*}
\resizebox{\hsize}{!}{
\includegraphics[]{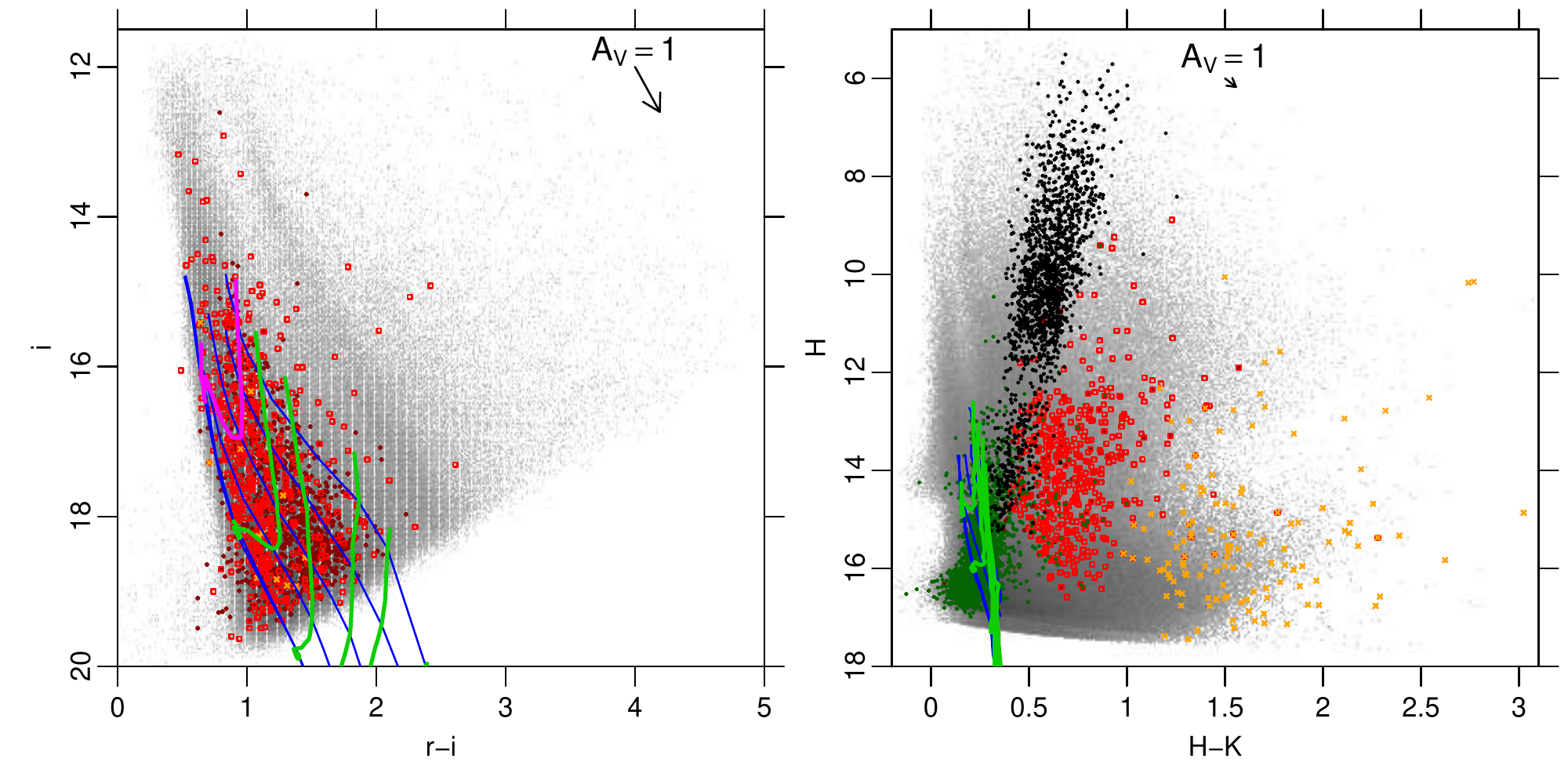}}
\caption{
Color-magnitude diagrams for all stars in the Sco~OB1 region (left:
$(i,r-i)$; right: $(H,H-K)$).
Only data with errors less than 0.1~mag on colors and magnitudes are shown.
Colored symbols, tracks and isochrones as in Figure~\ref{definitions}.
The magenta isochrone in the left panel corresponds to 1~$M_{\odot}$ stars.
{ In the $(H,H-K)$ diagram, dark-green dots indicate all candidate M
dwarf (or PMS) stars, while black dots are candidate M giants.}
\label{cmd}}
\end{figure*}

Figure~\ref{definitions}$f$ is instead the usual $(H-K,J-H)$ diagram,
from which, using the fiducial limit shown, that is:
\begin{equation}
(J-H) < -0.8+1.9\; (H-K)
\end{equation}
we select 233 NIR-excess stars. In the same diagram we show also the
optical-NIR excess stars, which follow well the CTTS locus predicted by
Meyer \e (1997), at low IR excess levels. Therefore, our
optical-NIR excess stars are unlikely to be dominated by spurious
identifications between stars in the optical and IR catalogs.

Finally, we consider stars with an UV excess, found using the VPHAS+ $u$
band. UV emission in PMS stars is commonly associated with an accretion
hot spot, where matter falling from the
circumstellar disk hits the star surface and heats up.
This type of measurement is often difficult both because the CCD
sensitivity at these wavelengths is low, and because UV radiation is
very sensitive to line-of-sight absorption toward the star.
The $(g-r,u-g)$ diagram of Figure~\ref{ug-gr} is an useful tool to
find UV-excess stars. As above we set a fiducial limit, above which
UV-excess stars are found (above $2\sigma$ significance), defined as:
\begin{equation}
(u-g) < -1.6+1.4\; (g-r)
\end{equation}
which yields 233 stars. Of them, 127 were already selected as \ha- or
NIR-excess stars. On the other hand, 132 stars with \ha- or
NIR excess and measured $u$ magnitude (with error less than 0.1~mag)
show no $u$-band excess.

{Stars with \ha-emission, optical-NIR excess, or $u$-band excess are
also reported in Table~\ref{table-mstars}.}

\subsection{IR-excess sources from Spitzer}
\label{spitzer}

Photometric data from the Spitzer space observatory are also very useful
to select populations of Class~II (or Class~0-I) stars, thanks to their
longer-wavelengths bands.
Glimpse observations cover most of the Sco~OB1 region studied here,
with the notable exception of NGC~6231 and immediate surroundings.
Figure~\ref{glimpse-col-col} is a $([5.8]-[8.0],[3.6]-[4.5])$
color-color diagram, which is effective to select Class~0-I-II sources
(Gutermuth \e 2009). We use this diagram here only to provide
(additional) evidence for youth of stars found across Sco~OB1, and
therefore we are here not interested in the fine details concerning
the IR classification of sources. Accordingly, we defined a threshold in
the $([5.8]-[8.0],[3.6]-[4.5])$ plane above which most Class~0-I-II sources
are expected to be found:
\begin{equation}
([3.6]-[4.5]) > 0.7 - ([5.8]-[8.0]);
\end{equation}
other types of sources (e.g., extragalactic
ones) are potentially found in the same region, but are unlikely to be
numerous in these low-latitude fields, and if any, they are not expected
to show spatial clustering, which is instead the case for PMS objects
{ (see Section~\ref{spatial})}.
The number of IR-excess objects selected in this way is 591.

\subsection{Color-magnitude diagrams}
\label{cmds}

\begin{figure*}
\resizebox{\hsize}{!}{
\includegraphics[angle=90]{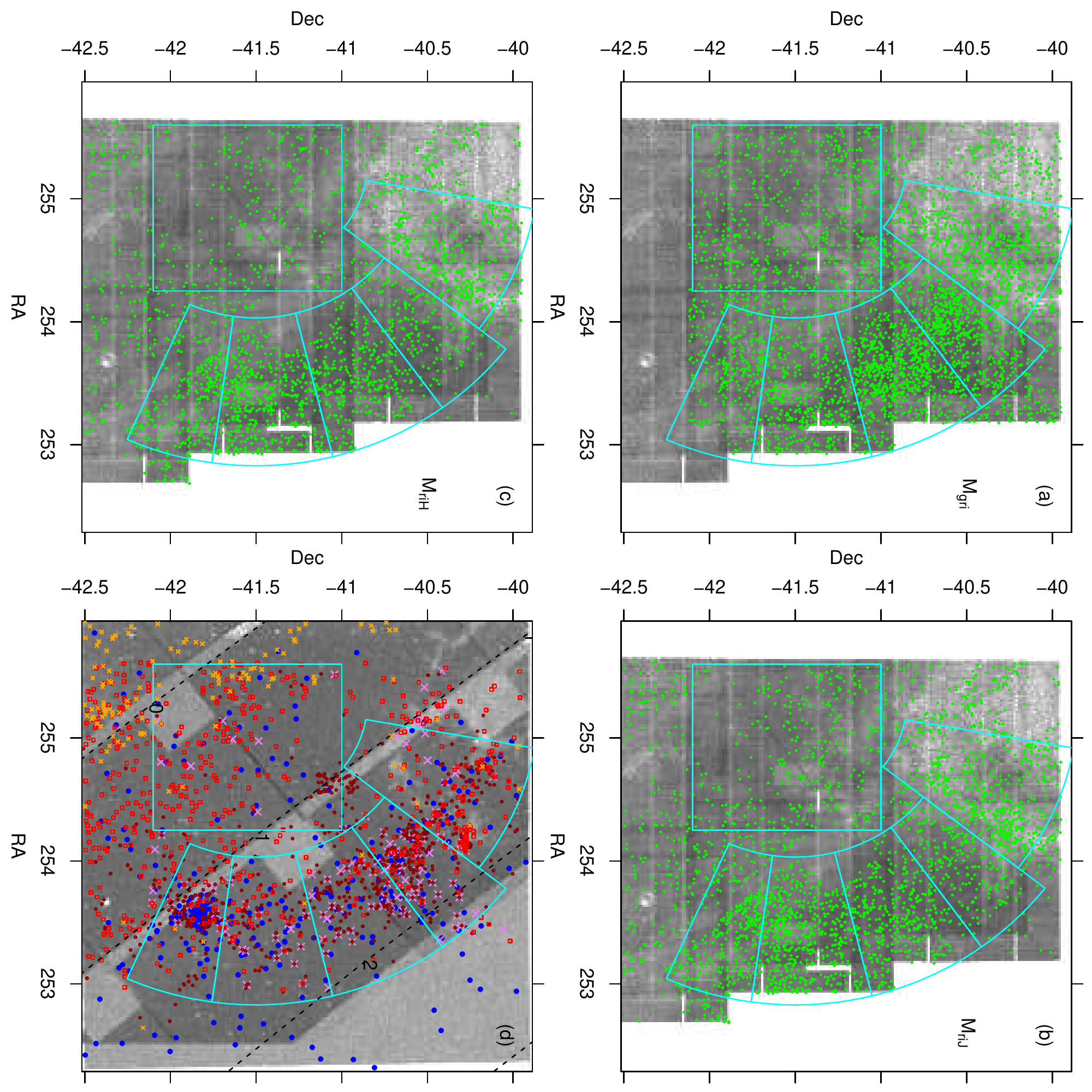}}
\caption{
Spatial distributions of star subsamples in Sco~OB1. The background
image is a 2-dimensional density histogram of stars in VPHAS+ DR2
(and sources in our NIR catalog in the bottom right panel).
Top left: $M_{gri}$ stars (green) selected from Fig.~\ref{definitions}$a$.
Top right: $M_{riJ}$ stars (green) from Fig.~\ref{definitions}$b$.
Bottom left: $M_{riH}$ stars (green) from Fig.~\ref{definitions}$c$.
Bottom right: dark-red points, red squares and orange crosses
are \ha-emission stars, and stars with optical-NIR and NIR excesses,
respectively, as in
Fig.~\ref{definitions}; larger purple crosses are UV-excess stars from
Fig.~\ref{ug-gr}; big blue dots are O or B stars from SIMBAD.
Oblique dashed black lines indicate constant galactic latitude $b$, as
labeled. Spatial subregions of interest are indicated in all panels with
cyan borders.
\label{spatial-Mstars}}
\end{figure*}

The optical and NIR color-magnitude diagrams (CMD) using our combined
optical-NIR catalog are shown in Figure~\ref{cmd}. Also shown are BHAC
evolutionary tracks and isochrones, for the reddening and distance
of NGC~6231.  In the $(i,r-i)$ diagram, the \ha-excess stars do not
form a narrow sequence, but are compatible with an age range between
approximately 1-10~Myr, in agreement with the findings by Damiani \e
(2016) for NGC~6231.  Many of the NIR-excess stars in the same diagram
lie in places suggesting older ages, contrary to expectations; however,
we show in Section~\ref{results} that the majority of these stars in
our region belong to a diffuse galactic-plane population, not to Sco~OB1.

The $(H,H-K)$ diagram in Figure~\ref{cmd} (right) shows that the bulk of NIR
sources in the region have very large extinction, of tens of magnitudes in $V$.
There is a remarkable change in the source density near $H \sim 14$,
which corresponds to the 2MASS-VVV transition: this is caused by incompleteness
in the VVV catalog after our filtering, as explained in Section~\ref{optnir}.
The red $H-K$ colors of the NIR-excess stars do not necessarily correspond
to high extinction, since they may largely be due to emission from
circumstellar disks.
{ The same Figure also shows clearly the magnitude difference between
our selected M dwarfs and giants:
as mentioned, despite their large reddening M giants are as a
class distinctly brighter than the selected M dwarfs, with a magnitude
gap around $H \sim 12.5$ between the two sets; a few stars in the ``giant''
subsample, fainter than that gap, might actually be dwarf stars.}

\section{Results: the Sco OB1 star-forming complex}
\label{results}

\subsection{Spatial morphology}
\label{spatial}

The {spatial distribution of the} stellar samples selected using the
methods described in Section~\ref{tech}
{ is} shown in Figure~\ref{spatial-Mstars}. The different panels permit to
understand in detail the spatial coverage of the VPHAS+ and VVV catalogs.
The VPHAS+ data cover uniformly our region, except for its western part.
The tiles used in the VVV survey do not cover seamlessly the region, as is
seen in the lower right panel: lighter background corresponds to a lower
NIR source density, where only 2MASS data are available.
The missing areas, however, are a minor fraction and do not critically
affect our results.
In the first three panels we also remark a reduced stellar density in the
VPHAS+ catalog in the northeastern corner, corresponding to the
region \gal, which is characterized by both diffuse nebulosity and
background obscuration. Moreover, the upper left panel show a clear lack of
$M_{gri}$ stars at $Dec<-42.1$, where $g$ band data are still missing in
VPHAS+ DR2.

The first three panels show separately the spatial distributions of $M_{gri}$,
$M_{riJ}$ and $M_{riH}$ stars, respectively. There is a nearly continuous
distribution of these stars spanning an arc-like region from NGC~6231 up to
\gal, with densities in excess of neighboring fields. We have drawn with a
large rectangle a reference field, where there is no indication of
young stars associated to Sco~OB1. Since the candidate PMS M stars are
found along a curved locus, we have instead selected five Sco~OB1 subregions,
also shown in the figure with sectors.
Our reasonings of Section~\ref{tech} indicate that the {
overdensities of}
M stars found in these sectors are {made of} good candidate members of this SFR.
The {lower-right} panel shows instead the spatial distributions of \ha-emission
stars, NIR- and UV-excess stars, and known OB stars: they describe a
{ locus remarkably similar to} the M stars, which strengthens our arguments
on the M stars membership to Sco~OB1.
In the lower left corner of the same panel, a large number of NIR-excess
sources is found, with density increasing toward the galactic plane.
These might be PMS stars or other dust-rich objects, but
{ their spatial distribution suggests that they are}
unrelated to Sco~OB1 and are not considered further.
Also, a small group of \ha-emission stars near $(RA,Dec)=(254.6,-41)$
have no counterpart among NIR-excess or other young stars: they also are
probably unrelated objects (or perhaps artifacts in the data).

{
The comparison between panels (a) to (c) of Figure~\ref{spatial-Mstars}
makes clear that the $M_{gri}$, $M_{riJ}$ and $M_{riH}$ star subsamples do
not mutually coincide (for instance, the $M_{gri}$ stars outnumber
$M_{riJ}$ stars in the second and third sector from top, but the reverse
is seen in the fourth sector), which is explained by differences in the
average properties of the stars (and their environments) among
different subregions.
This aspect is also discussed in Sect.~\ref{statistics} below.
The differences among sectors 3-5 from top are
likely explained as a result of decreasing foreground extinction from south to
north (supported also by the results of Sect.~\ref{extinction} and
Fig.~\ref{av-distrib}), where smaller extinction favors more detections
in the $g$ band. The second sector does not fit into the sequence,
however: here extinction is large, yet $M_{gri}$ stars are more numerous
than $M_{riJ}$ stars. This subregion contains the bright nebula IC4628
and hosts a rich population of \ha-emission, NIR-excess and UV-excess
stars (Fig.~\ref{spatial-Mstars}$d$), all indicating a very young age.
Therefore, the most likely explanation for the relative lack of $M_{riJ}$
and $M_{riH}$ stars here is that many M stars are still surrounded by
moderately massive disks, whose emission causes their $i-J$ and $i-H$
colors to appear redder than a (reddened) photosphere, and shifted
rightwards in the diagrams of Figure~\ref{definitions}$b$,$c$, by an
amount sufficient to escape our $M_{riJ}$ and $M_{riH}$ selection criteria
(though not necessarily as large as to cause these stars to be included
in the NIR-excess sample).
}

\begin{figure}
\resizebox{\hsize}{!}{
\includegraphics[angle=90]{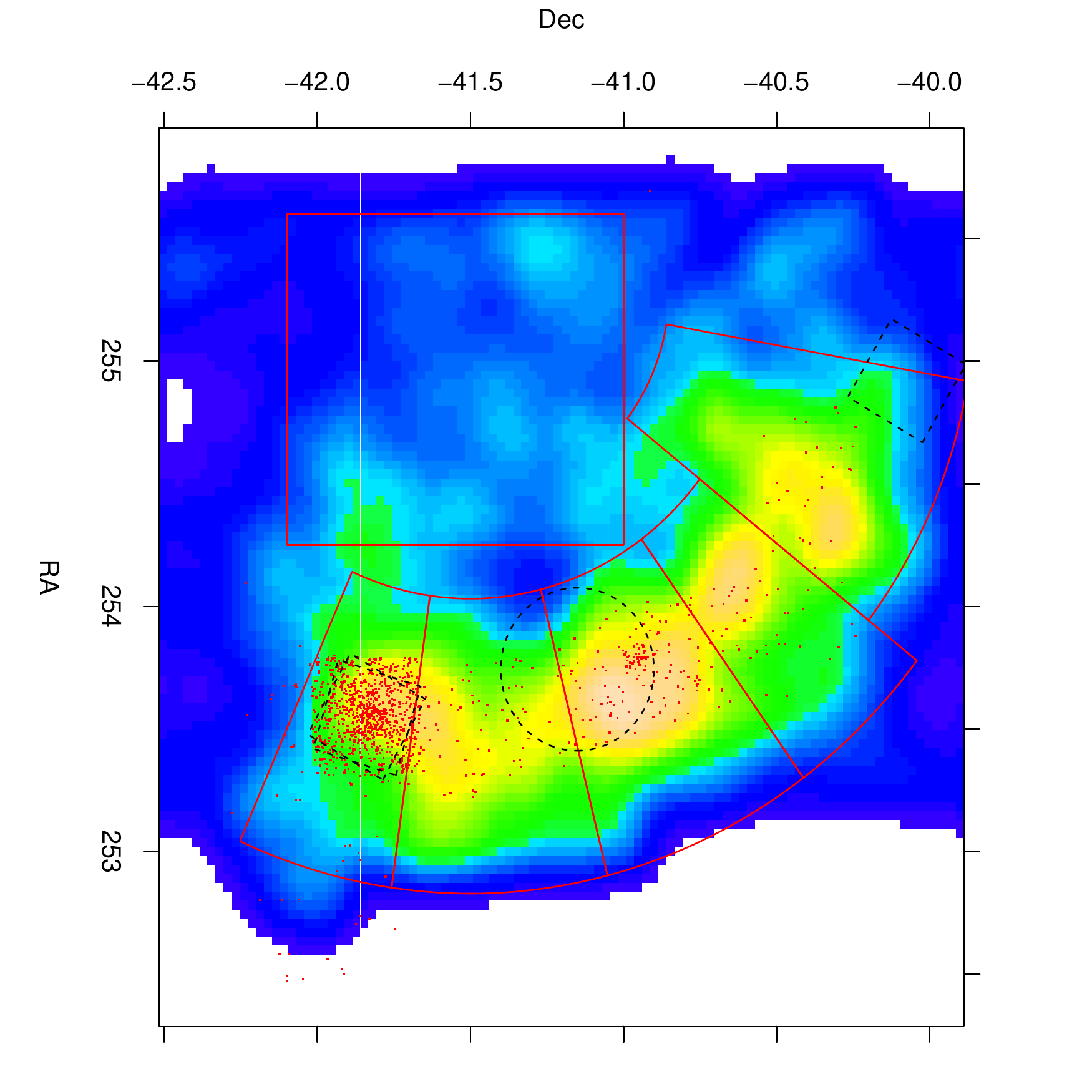}}
\caption{
Smoothed spatial distribution of all M stars (except for giants).
The black dashed regions are the same Chandra and XMM-Newton FOVs as in
Figure~\ref{dss-red}. Red regions are the same as in
Figure~\ref{spatial-Mstars}.
Red dots indicate all entries of type ``stars in cluster'' from the SIMBAD
database.
\label{smoothed-Mstars}}
\end{figure}

\begin{figure}
\resizebox{\hsize}{!}{
\includegraphics[width=8cm]{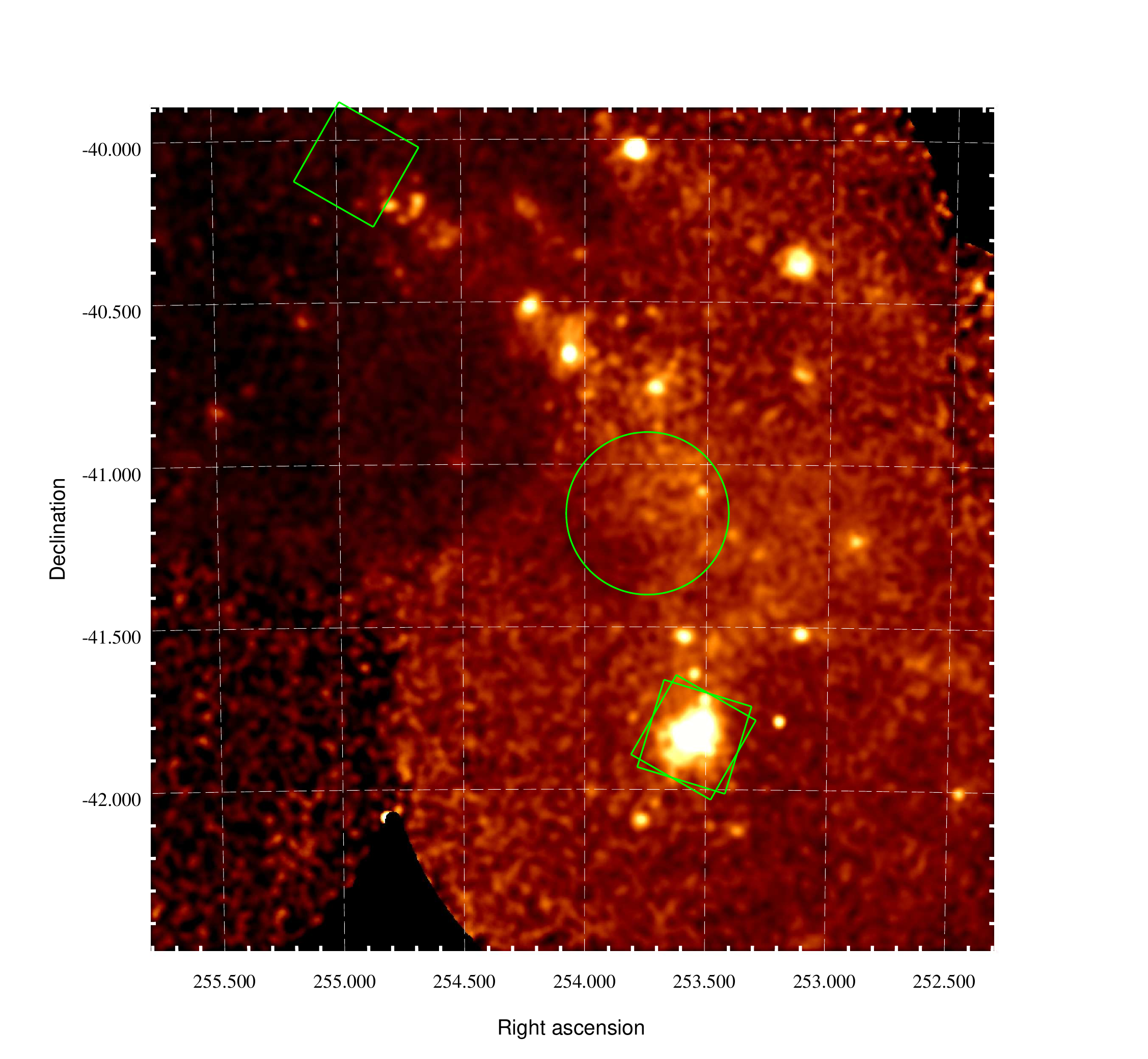}}
\caption{
ROSAT PSPC mosaic image of Sco~OB1, gaussian-smoothed with $\sigma=85$".
The green regions are the same Chandra and XMM-Newton FOVs as in
Figure~\ref{dss-red}.
\label{pspc}}
\end{figure}

A different representation of the spatial distribution of all M stars
we selected (by any method) is provided by Figure~\ref{smoothed-Mstars},
smoothed using a 5-arcmin Gaussian kernel
{(a 3-arcmin kernel does not result in a significantly different map)}.
{ Density variations shown by this map are not a byproduct of spatial
incompleteness: for example, the reduced density found in the $M_{riJ}$
and $M_{riH}$ maps at $RA \sim 253.5$, above the main arc
(Figure~\ref{spatial-Mstars}$b$ and~$c$), might be
ascribed to incompleteness in the NIR catalog, but a similar
underdensity pattern is seen in the locally complete $M_{gri}$ map
(Figure~\ref{spatial-Mstars}$a$);
similarly, the $M_{gri}$ map of Figure~\ref{spatial-Mstars}$a$ is
obviously affected by incompleteness for $Dec <-42$ and is not useful to
determine the southern bound of the arc-like region, but the $M_{riJ}$
map of Figure~\ref{spatial-Mstars}$b$ shows that there is no M-star overdensity
south of NGC~6231, so that the overall pattern shown by
Figure~\ref{smoothed-Mstars} can be considered reliable.
{The differences between subregions, discussed above, might to some
extent affect the relative heights of the density peaks in
Figure~\ref{smoothed-Mstars}, but much less their shapes, and since the
$M_{gri}$, $M_{riJ}$ and $M_{riH}$ selection methods tend to mutually
compensate their respective weaknesses, this map is much more complete than
any of the $M_{gri}$, $M_{riJ}$ or $M_{riH}$ maps alone.
}
The smoothed map of Figure~\ref{smoothed-Mstars} shows clearly} that the
M star distribution is not actually continuous but shows
{ four-five major clumps\footnote{ The densest part of the southernmost
clump coincides with the NGC~6231 cluster, but the smoothed M-star
distribution also shows a sort of tail to the north-west of NGC~6231,
which we prefer to consider as a group distinct from the cluster.}.}
Each of the sectors we defined {above corresponds to one such}
clump.  From North to South, they correspond to: (1) the \gal\ region;
(2) the IC4628 bright nebula, and the northern part of Tr~24;
(3) the central part of Tr~24; (4) the southern part of Tr~24; 
(5) the NGC~6231 cluster. The M star density outside of these five regions
is much lower. The \gal\ region has the least uniform distribution of M
stars, and also of NIR-excess sources (Figure~\ref{spatial-Mstars}, lower
right panel); therefore we extended its sector radially
toward the galactic plane.

{
We remark that NGC~6231, the densest and richest cluster in Sco~OB1,
does not correspond to the strongest peak in the M-star distribution in
Figure~\ref{smoothed-Mstars}. While this will be more quantitatively
discussed in Sect.~\ref{statistics} below, we may already explain the
reduced efficiency in M-star selection there from a combination of
higher extinction and older age (Sect.~\ref{extinction}
and~\ref{statistics}), coupled to our very conservative filtering of
input catalogs (Sect.~\ref{optnir}) which rejects comparatively more
stars in the densest regions, like the core of NGC~6231.
On the other hand, the \ha-emission and IR-excess stars shown in
Figure~\ref{spatial-Mstars}$d$ also include stars of types earlier than
M, and are less affected.
}

Figure~\ref{smoothed-Mstars} also shows the distribution of all stars
suspected to be members of some of the Sco~OB1 clusters, from the {
literature} studies
mentioned in Section~\ref{intro}, {which are of heterogeneous
nature}.
{
Noteworthy is the small group of stars, slightly offset to the East from
the Tr~24 density peak: this group is named C~1651-408 in SIMBAD, and
was also recently recovered in the study by Kuhn \e (2017b, their Figure~11).
It is somewhat puzzling that this small cluster is not seen among
\ha-emission or IR-excess stars (Figure~\ref{spatial-Mstars}$d$),
nor in X-rays (Figure~\ref{pspc} below), and therefore
increased extinction does not seem responsible for its non-detection as
a distinct peak in
the M-star sample; we conclude that this cluster is unlikely to be very
young, despite lying very close to Tr~24 on the sky.
}

In Figure~\ref{smoothed-Mstars} we also show the FOVs of existing X-ray
Chandra and XMM-Newton observations. However, the entire Sco~OB1 region was
also observed nearly two decades ago using the ROSAT X-ray satellite.
The detector used, the Position Sensitive Proportional Counter (PSPC),
had not the sensitivity nor the spatial resolution of Chandra or XMM-Newton,
but with respect to them had a much larger FOV ($2^{\circ}$ diameter).
The few existing PSPC pointings cover most of Sco~OB1, unlike the
Chandra and XMM-Newton data, and are shown in Figure~\ref{pspc}.
Most of the diffuse background is caused by energetic particles, not cosmic
X-rays, and varies between individual exposures. Nevertheless, local 
background structures {(a few arcmin in size)} are due to the
unresolved X-ray emission from hundreds of faint
sources; this {diffuse emission}, and a few individual,
brightest X-ray sources including
the entire cluster NGC~6231, describe an arc of enhanced emission
very similar to the M stars, as expected if these are indeed a large
population of X-ray bright low-mass PMS stars.

\begin{figure}
\resizebox{\hsize}{!}{
\includegraphics[]{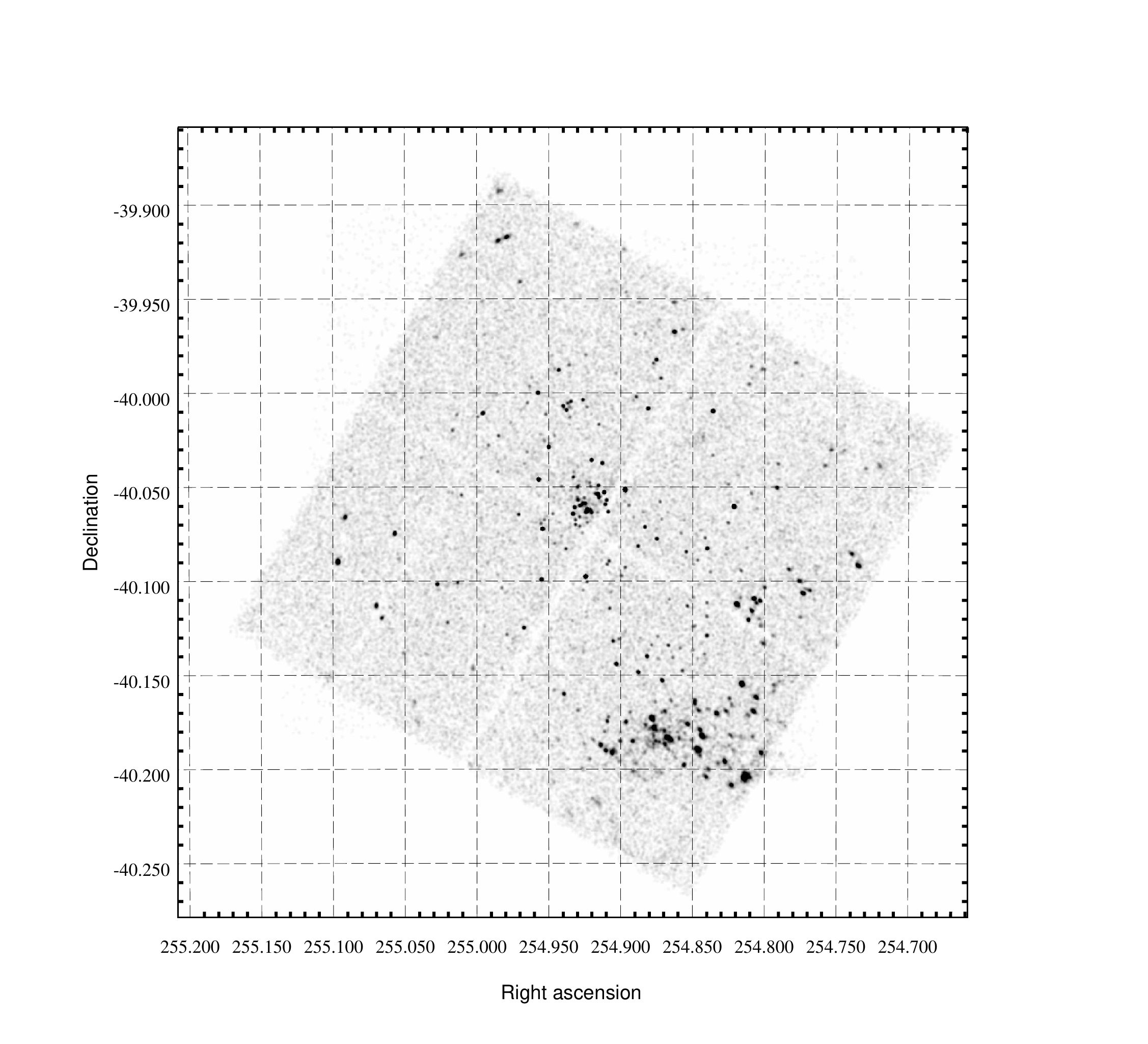}}
\caption{
Chandra ACIS-I image of the IRAS~16562-3959 region, slightly smoothed to
emphasize point sources.
\label{acis}}
\end{figure}

\begin{figure}
\resizebox{\hsize}{!}{
\includegraphics[angle=90]{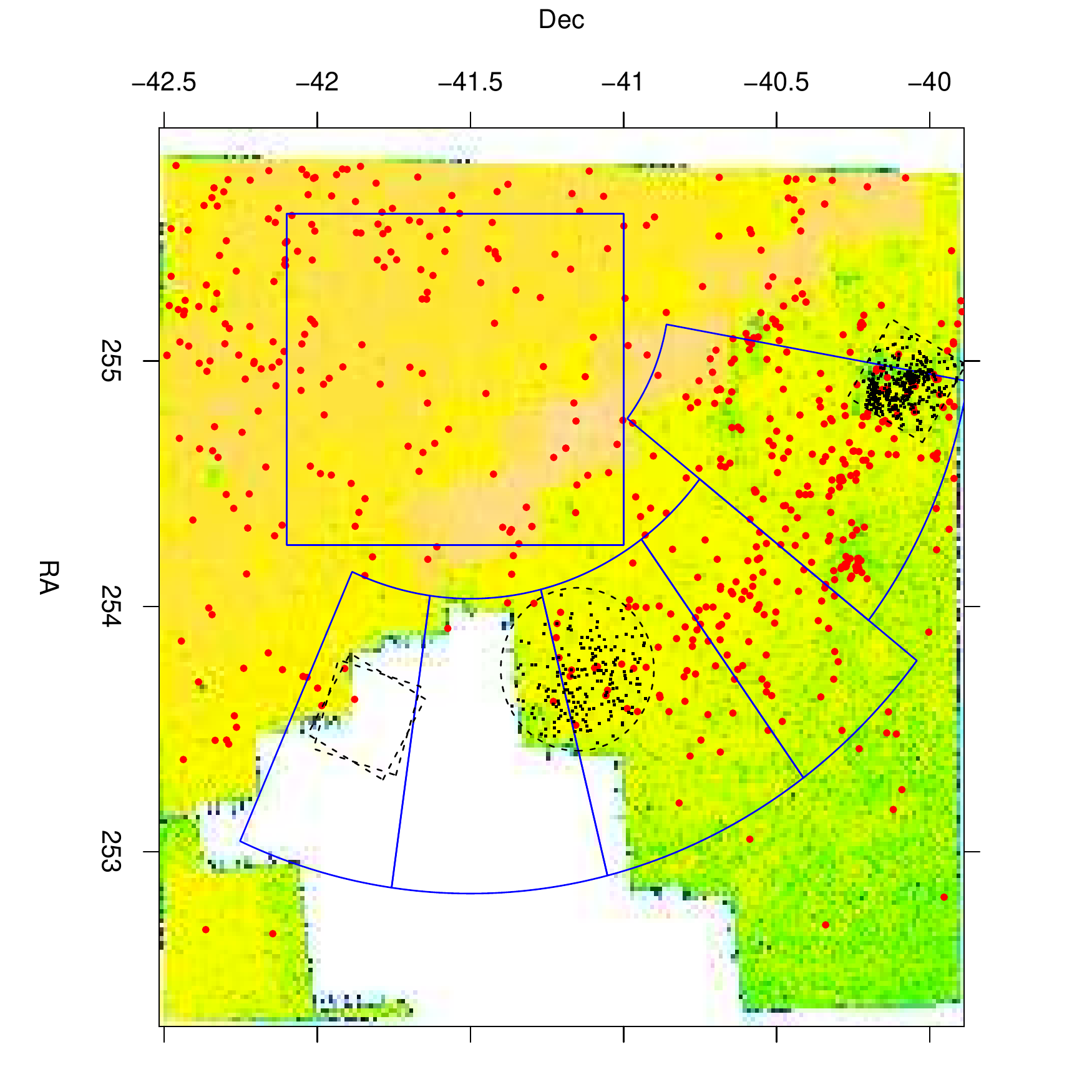}}
\caption{
Spatial-density 2-d histogram (shades of color) for NIR sources in the
Spitzer Glimpse catalog. Big red dots are Spitzer candidate YSOs. Small
black dots are X-ray detections. Chandra and XMM-Newton X-ray FOVs
(dashed black) are as in Figure~\ref{dss-red}.
Subregions (blue) as in Figure~\ref{spatial-Mstars}.
\label{spitzer-yso}}
\end{figure}

In the \gal\ region, the PSPC image shows enhanced emission inside the
southwestern part of the corresponding Chandra ACIS-I FOV, and it is
interesting to understand how the higher-quality Chandra data change
the picture there. The ACIS-I image (analyzed in Section~\ref{pwdet}) is
shown in Figure~\ref{acis}: the unresolved PSPC emission is now resolved into
{ a small cluster of}
more than 100 point X-ray sources; a distinct, smaller group of X-ray sources
is instead found surrounding the target IRAS~16562-3959.
{ These are most likely X-ray bright young stars inside the \gal\
\hii\ region.} Interestingly,
also the smoothed M-star distribution in Figure~\ref{smoothed-Mstars}
shows an enhanced M-star density in the southern part of the ACIS-I FOV.
This is one more confirmation of the PMS nature of our selected M stars.
Considering again the PSPC image, it also suggests an enhanced source
density in the northwestern part of the XMM-Newton image (not shown
for brevity); this is indeed the case, as shown by the distribution of
XMM-Newton X-ray detections in Figure~\ref{spitzer-yso}.
{ Therefore, the small clusters of X-ray sources seen in the
available X-ray images are in full agreement with the picture that our
M-star overdensities are indeed PMS members of Sco~OB1.}

\begin{figure}
\resizebox{\hsize}{!}{
\includegraphics[]{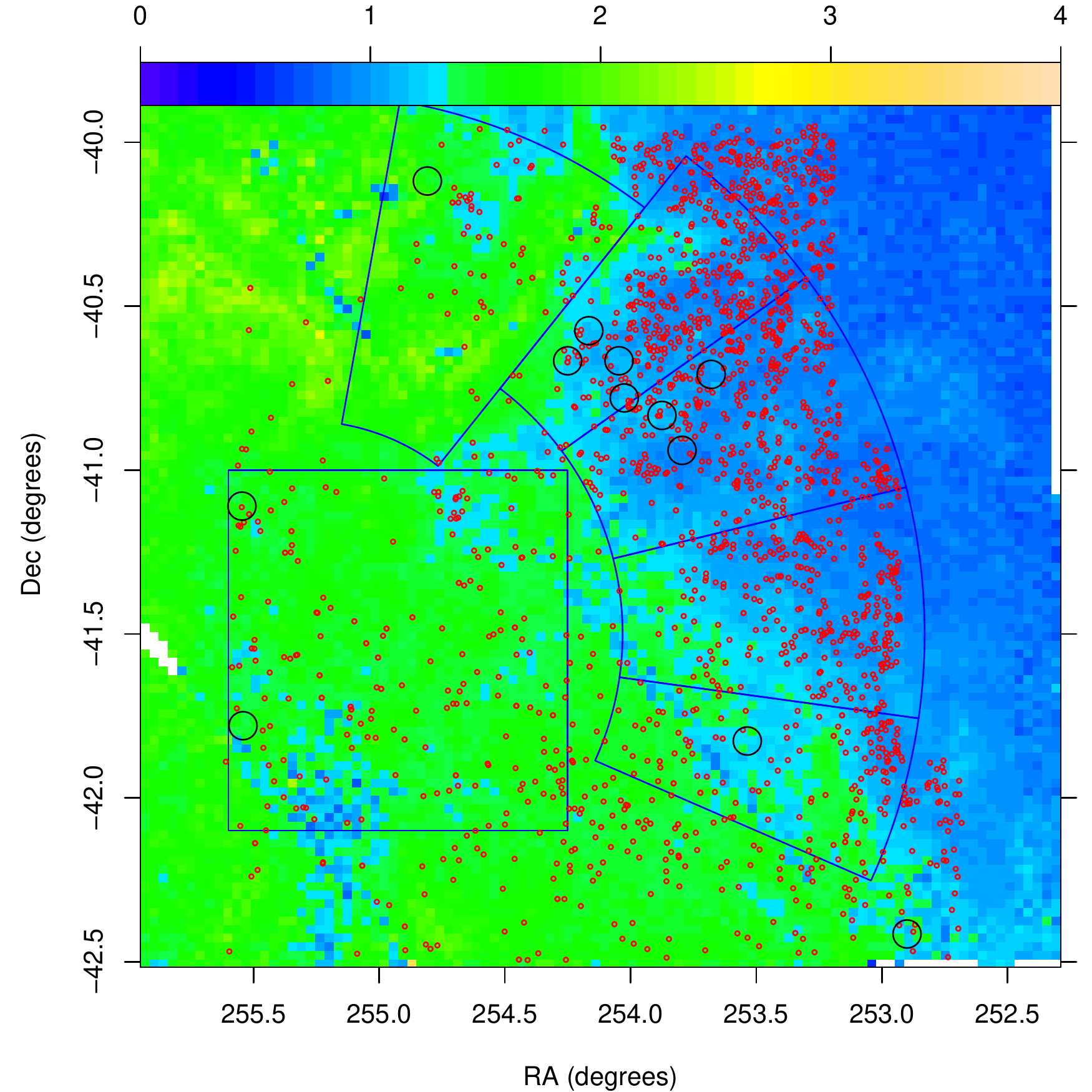}}
\caption{
Map of mean $(J-H)$ color (color scale shown in top axis).
Subregions (blue) are as in Figure~\ref{spatial-Mstars}.
Red circles are high-reddening M stars (giants) from
Figure~\ref{definitions}-$a,b,c$. Bigger black circles indicate
positions of known open clusters from SIMBAD.
\label{extinction-map}}
\end{figure}

Figure~\ref{spitzer-yso} {also} shows (as background) the spatial density of all
Spitzer IR sources in the region, from the Glimpse catalog. Unfortunately
the subregions Tr~24-South and NGC~6231 are not covered, which
prevents a complete study. The YSOs selected from Figure~\ref{glimpse-col-col}
are shown in Figure~\ref{spitzer-yso}: they are found in small groups
especially in the \gal\ and IC4628 regions, similarly to the NIR objects
from Figure~\ref{spatial-Mstars} and the X-ray sources in the
IRAS~16562-3959 Chandra field of Figure~\ref{acis}.
{ This is fully consistent with the extreme youth of \gal\ and
IC4628, while Tr~24 being older.}
The mentioned density enhancement of YSOs along the
galactic plane {(see Figure~\ref{wise})} is also found among Spitzer sources.
The Spitzer sources trace eminently galactic-disk objects, and their
density declines rapidly for increasing $b$ (that is, toward northwest).

In order to more fully understand the environment where the Sco~OB1 stars
are found (and were {likely} born), we have computed a map of mean $(J-H)$, which
is proportional to the average absorption toward objects in our
NIR catalog. This is shown in Figure~\ref{extinction-map}, where green color
corresponds to high-extinction regions (up to $A_V \sim 15$) and blue
to lower extinction. Some stratification parallel to the galactic 
plane is evident, but also remarkable is that the \gal\ region has
higher extinction than regions at the same {latitude} $b$. Tr~24 is projected
against lower-extinction regions, while NGC~6231 lies at intermediate
extinction values, and {was suggested to} have itself excavated a local hole in
the neighboring dust (Damiani \e 2016; also partially seen in
Figure~\ref{extinction-map}). The figure also shows the distribution of M
giants: these populate densely the low-extinction regions, where they
become observable up to large distances. Sco~OB1 being located
approximately $16^{\circ}$ from the galactic center, these M giants belong
probably to the galactic bulge. Therefore, their measured extinction provides
a strong upper bound to the {foreground} dust column density toward the studied region.
In the same figure we also show the positions of all {known} open clusters from
SIMBAD: apart from NGC~6231 and Tr~24, the other Sco~OB1 clusters
are very poorly known,
small clusters (ESO~332-11, C~1652-405, ESO~332-13 in the IC4628 region;
C~1651-408, C~1652-407, ESO~332-8 in the Tr~24 region). Interestingly,
in the \gal\ region the only IR-discovered cluster is found ([DBS2003]
115, Dutra \e 2003), which is also coincident with the southernmost
X-ray source cluster clearly seen in the Chandra image of
Figure~\ref{acis}, {once again confirming its embedded nature and
young age}.
{ The apparent spatial distribution of M giants, so tightly related to
the total line-of-sight absorption, is markedly different from
that of our candidate PMS M stars, which rules out that spatial
overdensities of the latter are caused by holes in the dust column
density.}

\begin{figure}
\resizebox{\hsize}{!}{
\includegraphics[]{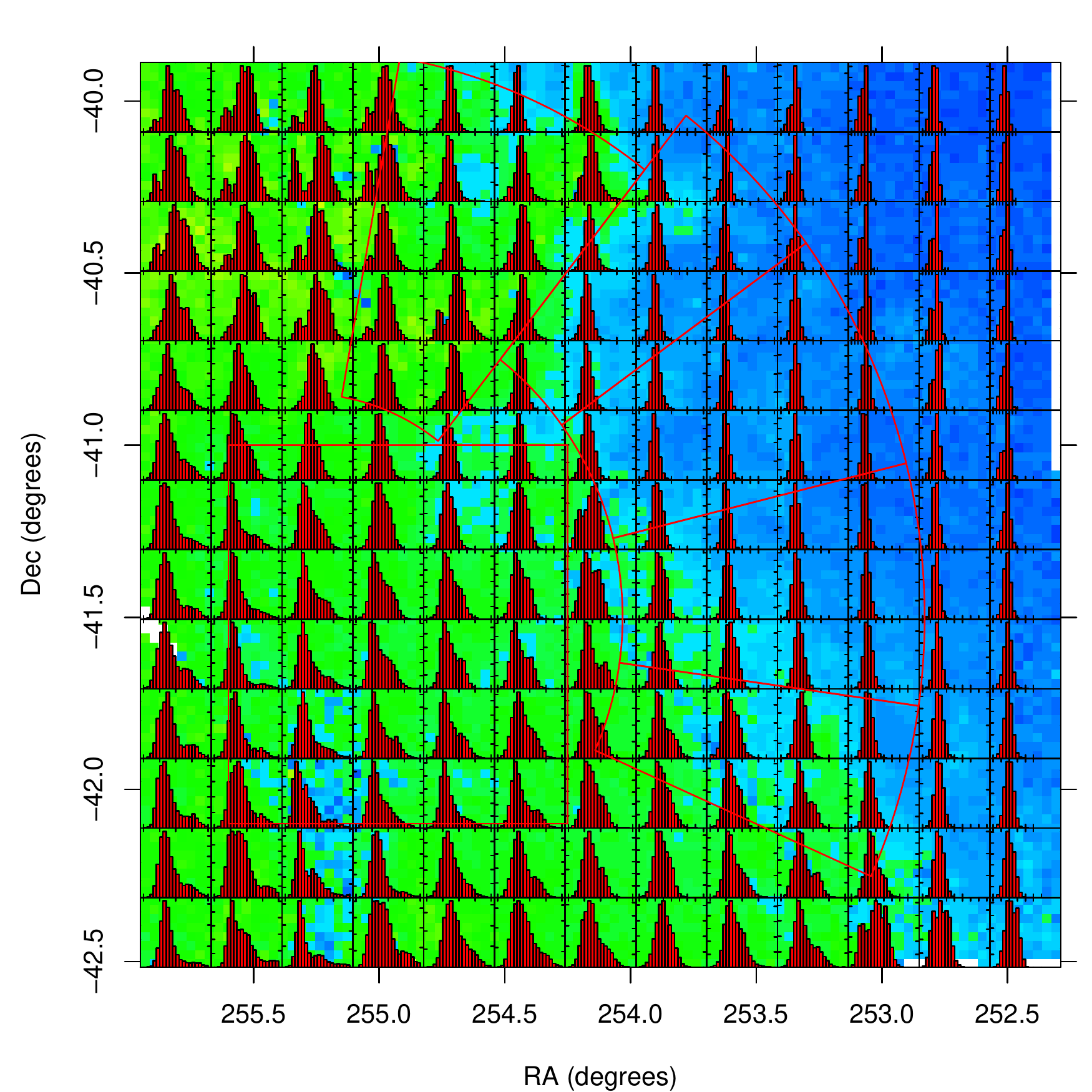}}
\caption{
Local distribution of reddening, from $(J-H)$ histograms (insets). The
background image is the same as in Figure~\ref{extinction-map}.
Subregions (red) are as in Figure~\ref{spatial-Mstars}.
\label{jh-hist}}
\end{figure}

Since absorbing dust is likely distributed over a wide range of distances
along any given sightline, the average extinction obtained from $(J-H)$
provides only a rough description, being simply the first-order moment
of the actual distribution in a given direction.
A more detailed knowledge of the dust distribution is obtained by considering
the histogram of $(J-H)$, in localized sky regions.
The result of this approach can be seen in Figure~\ref{jh-hist}, where each
histogram refers to a square spatial region of side $12^{\prime}$.
Each histogram covers a $(J-H)$ range of $[0-4]$, corresponding to a maximum
$A_V \sim 35$ magnitudes. The higher-latitude regions (with blue background)
are confirmed to show narrow $(J-H)$ distributions, with no tails at large
values: this confirms that the total (not only the average) extinction in
these directions is low enough $(A_V \leq 10)$ as to allow detection of
bulge M giants. On the other hand, at lower latitudes tails at high $(J-H)$
are regularly found.       
In the \gal\ region, the $(J-H)$ histogram is most complex, with two strong
and well-separated peaks, and significant variations over adjacent spatial
bins. The double peaks indicate a rapid rise of extinction with
distance,
{ caused by} a thick dust layer or cloud, so that stars in its immediate foreground and
background have markedly different $(J-H)$ colors, respectively,
while $(J-H)$ variations are smoother along the rest of the line of sight.
The Tr~24 regions (central, northern and southern) are among the most
transparent; this suggests that no residual dust from the epoch of
its formation has remained close to the cluster stars.

\subsection{Extinction to M stars}
\label{extinction}

\begin{figure}
\resizebox{\hsize}{!}{
\includegraphics[]{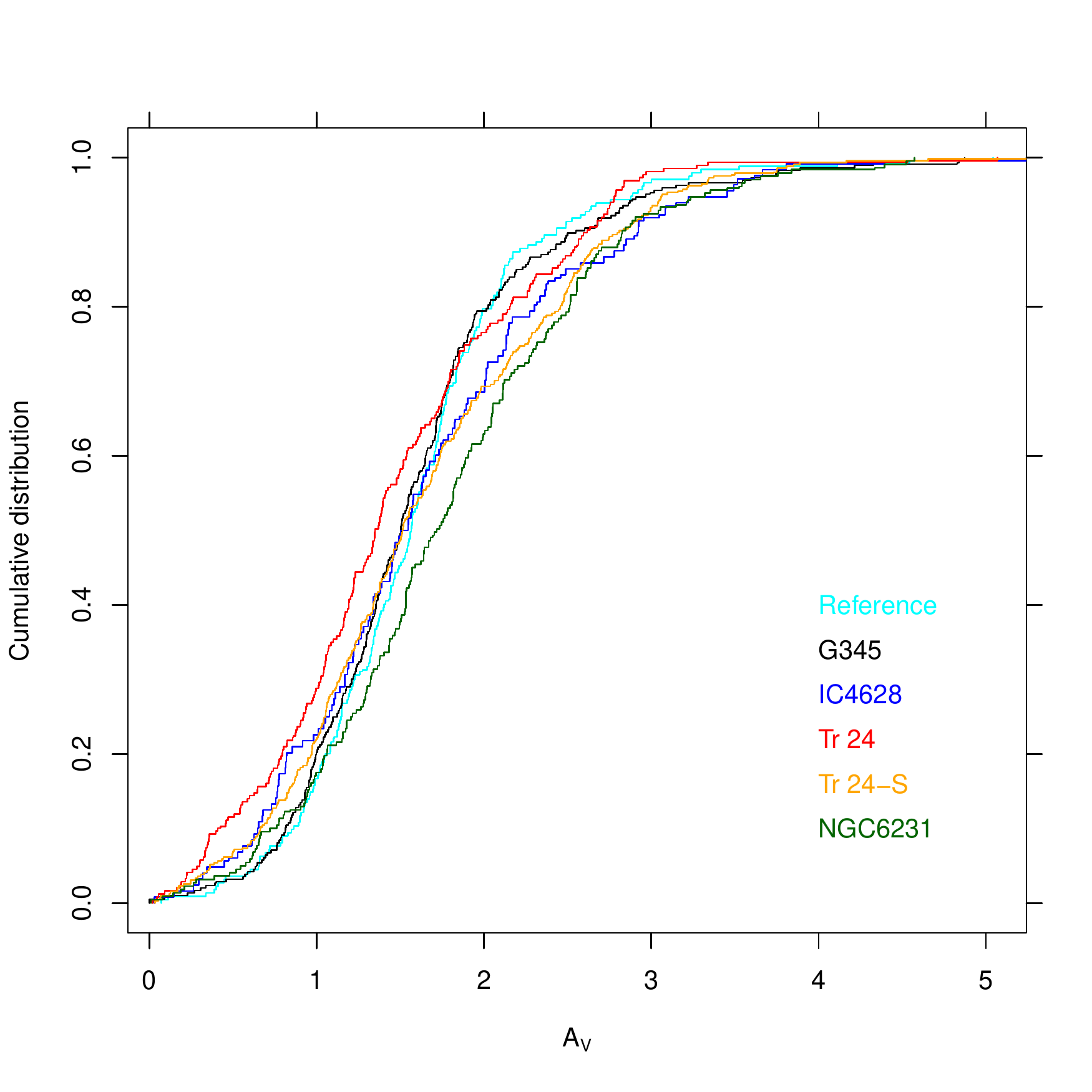}}
\caption{
Cumulative distribution of $A_V$ for subregions, including the reference
field.
\label{av-distrib}}
\end{figure}

As we mentioned in Section~\ref{tech}, the colors of M stars are
not degenerate with respect to extinction. This offers the opportunity
for measuring extinction $A_V$ for individual stars. The accuracy of $A_V$
determinations depends on uncertainties both on the photometry itself and on the
theoretical isochrones. At the level of precision of the available
optical and NIR photometry, differences between the two isochrone sets
used (Siess \e 2000 and BHAC) are significant, and $A_V$ derived from
the $(g-r,r-i)$ diagram and Siess models (the BHAC models being
unavailable for the $g$ band) is only roughly correlated with $A_V$ derived
from the $(i-J,r-i)$ diagram and BHAC models. Since the BHAC models appear
to reproduce better the observed datapoints in the $(i-J,r-i)$ diagram
of Figure~\ref{definitions}, and also the CMD of NGC~6231 X-ray members
in Damiani \e (2016), we consider in the following only $A_V$ obtained from
the $(i-J,r-i)$ diagram and BHAC models.

Figure~\ref{av-distrib} shows cumulative $A_V$ distributions for M stars
(not giants) in each of the five Sco~OB1 subregions (and reference region)
defined above.
Values of $A_V<0$ were reset to zero when compatible with it within errors.
Stars in NGC~6231 are characterized by the highest $A_V$ distribution,
which may be related to its relatively low latitude $b$ compared to the
other subregions. Schild \e (1971) and Neckel and Klare (1980) find that
most of the extinction
toward this cluster arises much closer to us along the line of sight.
Therefore, the enhanced $A_V$ toward this cluster does not necessarily imply
that its neighborhood contains more dust than the other subregions.
On the other hand, the \gal\ region, which is an active star-forming region
as summarized in Section~\ref{intro}, and contains a thick dust cloud as seen
from Figure~\ref{jh-hist}, has only M stars with relatively low $A_V$:
these are probably found on its near side, closer to us than the dusty
layers.
We compared the $A_V$ distributions {pairwise} between subregions using
Kolmogorov-Smirnov tests. All visual differences in
Figure~\ref{av-distrib} are statistically very significant: for example,
the NGC~6231 extinction is larger than that in the neighboring Tr~24-S
region at the 99.97\% confidence level. This latter, on the other hand, is
larger than that in the central Tr~24 region at the 99.9\% level.
The central Tr~24 region has the lowest extinction, both foreground as seen
from Figure~\ref{av-distrib}, and in its background as seen from
Figure~\ref{jh-hist}.

\subsection{Member statistics and ages of M stars}
\label{statistics}

\begin{figure}
\resizebox{\hsize}{!}{
\includegraphics[]{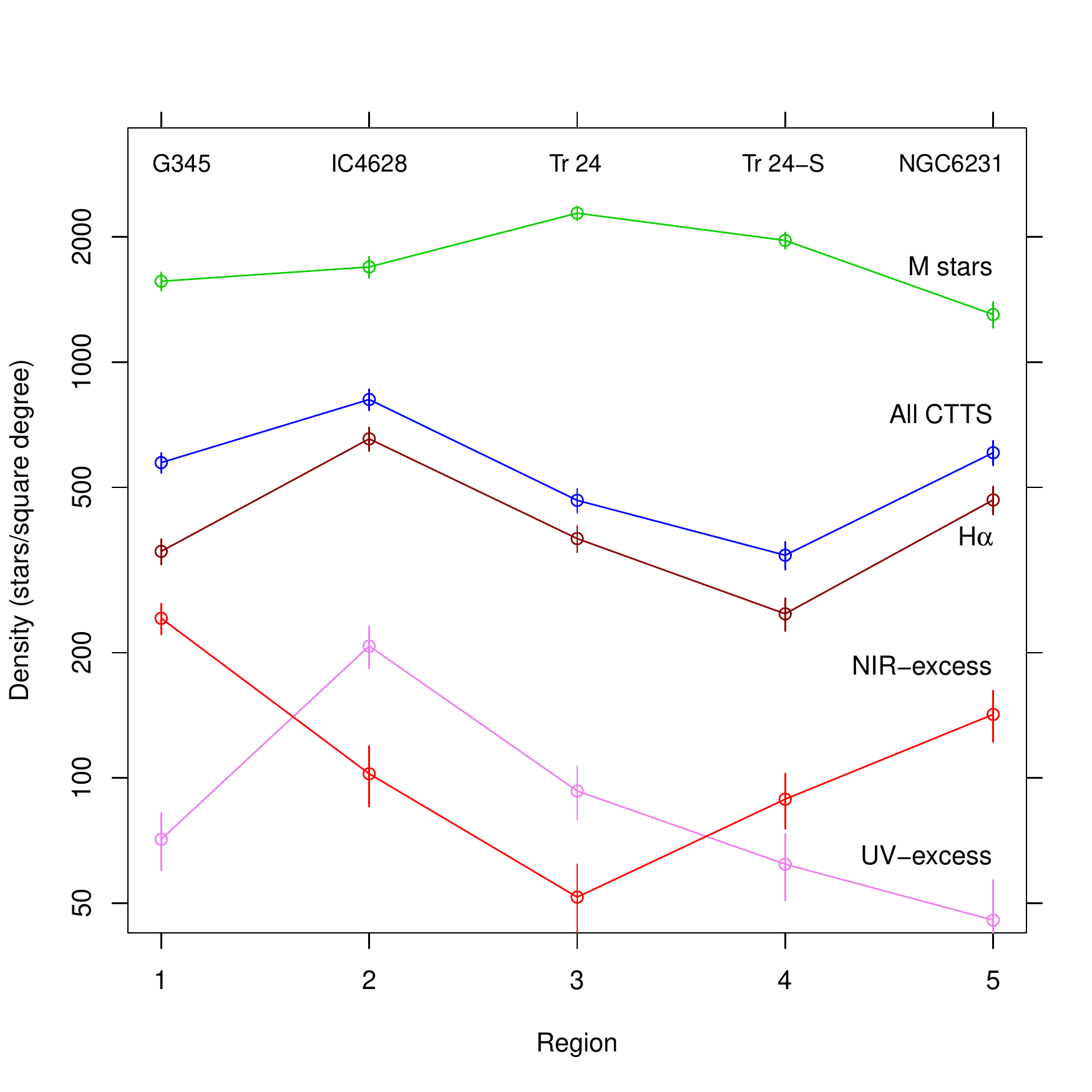}}
\caption{
Spatial density of M stars, \ha-excess, NIR-excess, and UV-excess stars
for the different regions.
\label{dens-regions}}
\end{figure}

\begin{table*}[ht]
{\small
{\bf Table 4.}
M stars and young objects in Sco~OB1 subregions.
Column Area is the region area in square degrees.
Column $M_{tot}$ is the total number of M stars.
Column Net reports the net number of M stars.
Column Density is the net M-star density in units of stars/sq.deg..
Column $f_C$ gives the number ratio between M non-members and all M stars
in the given subregion.
Columns \ha, NIR, UV, and OB report numbers of \ha\ emission, NIR-excess,
UV-excess, and (net) OB stars, respectively.
\label{table-stat}
}\\

\begin{tabular}{rlrrrrrrrrrrrrrr}
  \hline
 N & Region & Area & $M_{gri}$ & $M_{riJ}$ & $M_{riH}$ & $M_{tot}$ &
Net & Net & Density & Density & $f_C$ & \ha\ & NIR & UV & OB \\
  & name & & & & & & & Err. & & Err. & & & & & \\
  \hline
  0 & Reference & 1.11 & 939 & 604 & 458 & 1502 & 0.00 & 54.81 & 0.0 & 49.3 & 1.00 &  65 & 221 &  12 &  \\ 
    1 & G345 & 0.56 & 969 & 856 & 668 & 1638 & 879.04 & 44.96 & 1564.1 & 80.0 & 0.46 & 197 & 136 &  40 & 8 \\ 
    2 & IC4628 & 0.35 & 732 & 348 & 290 & 1071 & 595.82 & 34.95 & 1693.4 & 99.3 & 0.44 & 230 &  36 &  73 & 32 \\ 
    3 & Tr 24 & 0.48 & 1056 & 697 & 625 & 1757 & 1103.63 & 45.18 & 2281.1 & 93.4 & 0.37 & 182 &  25 &  45 & 26 \\ 
    4 & Tr 24-S & 0.48 & 593 & 1015 & 837 & 1602 & 948.63 & 43.43 & 1960.8 & 89.8 & 0.41 & 120 &  43 &  30 & 22 \\ 
    5 & NGC6231 & 0.35 & 340 & 598 & 480 & 933 & 457.82 & 32.91 & 1301.2 & 93.5 & 0.51 & 164 &  50 &  16 & 94 \\ 
   \hline
\end{tabular}
\end{table*}

While we argued in Section~\ref{spatial} above the qualitative evidence that
most selected M stars are indeed tracing the low-mass population of
Sco~OB1, we here discuss their quantitative statistics, and compare it with
{ member samples found using other techniques}.
Table~4 (and Figure~\ref{dens-regions})
reports the number of M stars
(by type and cumulatively) in each subregion, together with their
estimated net number, obtained by subtracting the (area-scaled) number
of stars in the reference field. Also given are the {M-star net} mean
surface densities,
and the estimated fractions $f_C$ of non-member contaminants, together
with the numbers of member stars selected using the other techniques
discussed in section~\ref{tech}.
{
Although the absolute numbers of \ha-emission and IR-excess stars in
NGC~6231 are among the largest of all subregions, they become the
smallest in relative terms, once they are scaled to
the respective numbers of OB stars. This will be discussed more deeply
below in connection with relative cluster ages.
}
{
Contaminant fractions $f_C$ are computed using the detected M-star
population in the Reference field, with an area scaling factor for each
sector. As discussed in Sect.~\ref{mstars}, these contaminants are
mostly main-sequence M stars, with a negligible fraction of low-reddening
giants. This procedure relies on the assumption that the distribution of
absorbing dust with distance from us is approximately the same in the
Reference field and in the Sco~OB1 subregions, which is to a first
approximation true from consideration of Figure~\ref{av-distrib}
(as far as only M stars are concerned).
}
{We note {from Table~4}
that the level of contamination is significant ($f_C \sim
0.37-0.51$), and therefore the membership of any particular star in our
sample can only be assessed using additional data; however, the
(relative) error on the net number of stars is so small to imply a
highly statistically significant population of Sco~OB1 members,
in any of the considered subregions.}

{
Unlike M stars, there are virtually no \ha-emission nor IR-excess
stars just outside Sco~OB1 subregions.  The Reference field does
contain many of them, but it was argued in Sect.~\ref{spatial} that
they are populations of stars lying at very low galactic latitudes, not
extending at the slightly higher latitude of Sco~OB1. The contamination
of \ha-emission and IR-excess Sco~OB1 stars can therefore be considered
negligible.
}

We note that there are large differences in the $M_{gri}/M_{riJ}$ number
ratio between different subregions, which might be at least partially
related to differences in foreground extinction (the blue $g$ band being
more affected than the $J$ NIR band), and by consequence different
minimum detectable stellar mass, from {the MDA diagrams of} Figure~\ref{max-distance}.
This latter figure, however, predicts that the minimum detectable mass
using the $(r,i,J)$ bands is always lower (using this particular photometric
dataset) that the minimum detectable mass using the $(g,r,i)$ bands;
therefore, the number of $M_{gri}$ stars should always be lower than that
of $M_{riJ}$ stars.
This does not appear to be the case in the IC4628 or Tr~24 subregions, where
$M_{gri}$ stars evidently outnumber $M_{riJ}$ stars. This peculiarity may be
explained by recalling that the adopted filtering for the VVV data
implied a (confusion-related) incompleteness by almost a factor of three
in the NIR catalog (see Section~\ref{data}).  The numbers of $M_{riJ}$
and $M_{riH}$ stars found are therefore to be regarded as underestimates,
by a factor {up to $\sim 2$.
{As mentioned in Section~\ref{spatial}, also circumstellar disk
emission may deplete the $M_{riJ}$ and $M_{riH}$ samples relative to the
$M_{gri}$ sample.}
The simultaneous use of several triplets of bands proves therefore
useful in that they complement each other, such that incompleteness in the
final selected population is reduced to a minimum.}

\begin{figure}
\resizebox{\hsize}{!}{
\includegraphics[]{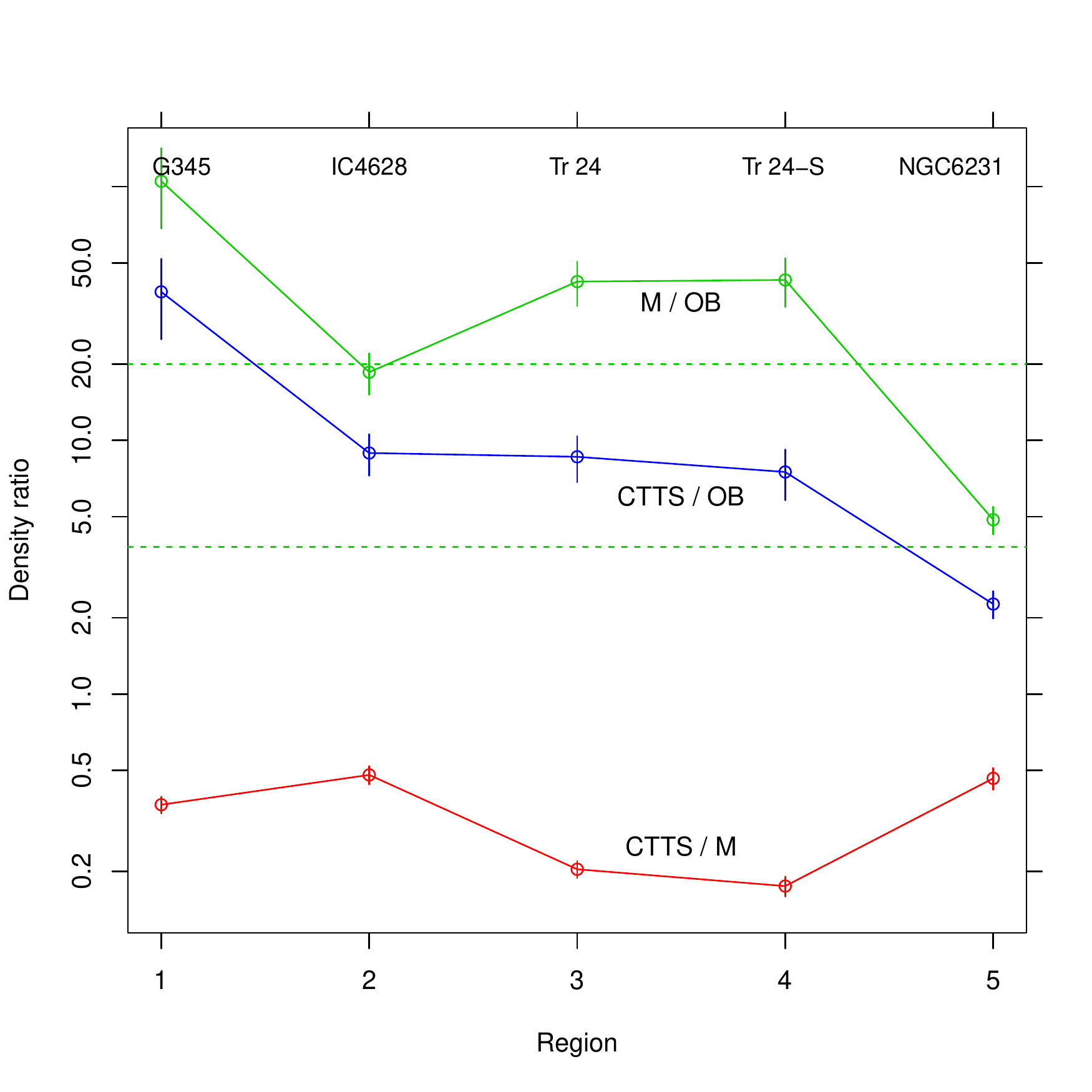}}
\caption{Net stellar density ratios between M and OB stars, cumulative
CTTS and OB stars, and CTTS and M stars, for each subregion.
The green dashed lines indicate the maximum and minimum values expected
for the M/OB number ratio.
\label{ctts-ob}}
\end{figure}

{ The last column in Table~4
reports the number of OB
stars, net of the ``diffuse'' population estimated from their density in
the Reference field (22.5 stars per square degree): as is immediately
seen, the M-star statistics is much larger than the OB star statistics.
This can hardly be considered surprising, if an ordinary IMF (e.g.\ that
from Weidner \e 2010) is assumed for the SFR.}
{ In Figure~\ref{ctts-ob} we show the density ratio between M
and OB stars, which provides a consistency test between our results
and a plausible IMF: this ratio varies however by a large factor, close
to 20, among our subregions.
This might reflect differences in the respective IMFs, but also
differences in completeness among the stellar samples considered for the
various regions.
We first note that the ratio between M and OB stars in NGC~6231 is dramatically
lower than anywhere else in Sco~OB1.
We can indeed expect that M stars are detected
less efficiently in the inner parts of NGC~6231, where the density of bright
stars is very large, and their diffuse glare raises the limiting
magnitude locally.
{As already discussed above in Sect.~\ref{spatial}, this causes our
sample of M stars in NGC~6231 to be highly incomplete.}
Moreover,
we have determined above that NGC~6231 is significantly more extincted,
by almost half a magnitude in $V$, than Tr~24, and this implies a higher
minimum detectable mass among NGC~6231 M stars, compared to Tr~24
(see {the MDA diagrams} in Figure~\ref{max-distance}); this effect
reduces the completeness of the M-star sample in NGC~6231 more
than in Tr~24.
If Tr~24 is also slightly younger than NGC~6231, as we argue below,
our M-stars in Tr~24 will reach down to lower masses than in
NGC~6231, with a steep increase in the detected M-star population:
adopting the IMF from Weidner \e (2010), the predicted number of
cluster M stars doubles considering the mass interval $0.25-0.5 M_{\odot}$
rather than $0.35-0.5 M_{\odot}$.
If Tr~24 is younger than NGC~6231, moreover, its stars in the mass range
$2.5-3 M_{\odot}$ might not have yet reached their ZAMS position as B stars,
and therefore would not be counted among OB stars; this would further raise
the M/OB star ratio there by up to 30\%.
}
Therefore, the proportions of both M and OB stars that are
detected in a young cluster will depend on their age and extinction, {in
accordance with the MDA diagrams, even for a fixed, spatially uniform
photometric sensitivity}. We
estimated using the Weidner \e (2010) IMF the expected range for the
{ observed M/OB} number ratio. Siess \e (2000) predict that the latest-type B stars
have a mass of $\sim 3.5 M_{\odot}$ at 2~Myr, and $\sim 2.2 M_{\odot}$
at 10~Myr, that is in the range of ages expected for Sco~OB1 clusters.
{ The MDA diagrams of} Figure~\ref{max-distance} predict that the lowest-mass stars we are
able to detect using the available Sco-OB1 data have $\sim 0.2
M_{\odot}$, even assuming the most favorable (and unlikely)
circumstances of an age less than 2~Myr and negligible reddening.
The extreme values found for the M/OB ratio are then $\sim 3.8$ for
a minimum M-star mass as high as $0.35 M_{\odot}$ and an old age of
10~Myr, and $\sim 20$ for a minimum M-star mass as low as $0.2
M_{\odot}$ and age of 2~Myr. These extremes are also shown {as
horizontal lines} in Figure~\ref{ctts-ob}.
We remark that the M/OB ratio in NGC~6231 falls well within this range;
however, both Tr~24 regions are significantly richer of M stars than
expected, by more than a factor of 2 and well above (statistical)
errors. If true, then paradoxically this part of the OB association
would form preferentially lower mass stars. Of course, more detailed
studies will be needed to confirm this result.
In the \gal\ region the M/OB ratio is highest, {and far above
predictions from the IMF}: we may tentatively
explain this since this region is very young, and some of its most massive
members, like IRAS~16562-3959, are still in formation, {thus
decreasing the number of optically revealed OB stars}.
{
The lowest M/OB ratio in NGC~6231 is unlikely to be real, since as
discussed above our M-star sample in this densest subregion is likely
incomplete.
}

{ We examine here if an age sequence can be identified among these
subgroups. As explained in Section~\ref{intro}, \gal\ is surely a very
young region (it contains masers, bright and dark nebulosity, and its stars
show an irregular spatial distribution), while on the other hand at the
opposite extreme NGC~6231 is much older (it contains no protostars nor stars
younger than
1~Myr, and has a relaxed morphology). The question is therefore if the
ages of the other subgroups lie along an ordered sequence, of which \gal\
and NGC~6231 are the extremes.}

{ Figure~\ref{dens-regions} shows that}
in proceeding from Tr~24 toward \gal, the density of NIR-excess stars
increases, as expected {if an age sequence exists}.
However, both the \ha-emission and UV-excess stars
decrease in density, contrary to expectations. This may be caused by 
locally increased \ha\ background in \gal, and by circumstellar absorption
near the youngest stars, which obstacle their detection
{ in the \ha\ and $u$ bands, respectively. As a result, no single
indicator among \ha\ emission, NIR or UV excess can be taken as a
reliable indicator of youth here}.

We {instead} consider collectively all PMS stars with NIR excess (Class~II), \ha\
emission or UV excess as CTTS (also shown in Figure~\ref{dens-regions}).
The number ratio between CTTS and the total population in a subregion may
be taken as {a better} indicator of its youth.
We have computed density ratios between
CTTS and M stars, and between CTTS and OB stars
(see Figure~\ref{ctts-ob});
{ Since as shown above the M/OB density ratio varies wildly between
subregions, the CTTS/M and CTTS/OB ratios do not show the same trend
across regions. The CTTS/M ratio does not seem to be a reliable
indicator of age, being essentially the same in \gal\ and NGC~6231,
which are of different age with high certainty. The CTTS/OB ratio, on the
other hand, is fully consistent with our expectations of an increased
proportion of CTTS stars in \gal\ than anywhere else in Sco~OB1, and
shows a regular monotonic increase from NGC~6231 to \gal.}
This suggests a sequence of decreasing age, and {thus} that star
formation has progressed in a {regular fashion}
from the oldest NGC~6231 region, through the intermediate age Tr~24 and
up to the youngest region \gal.

\begin{figure}
\resizebox{\hsize}{!}{
\includegraphics[]{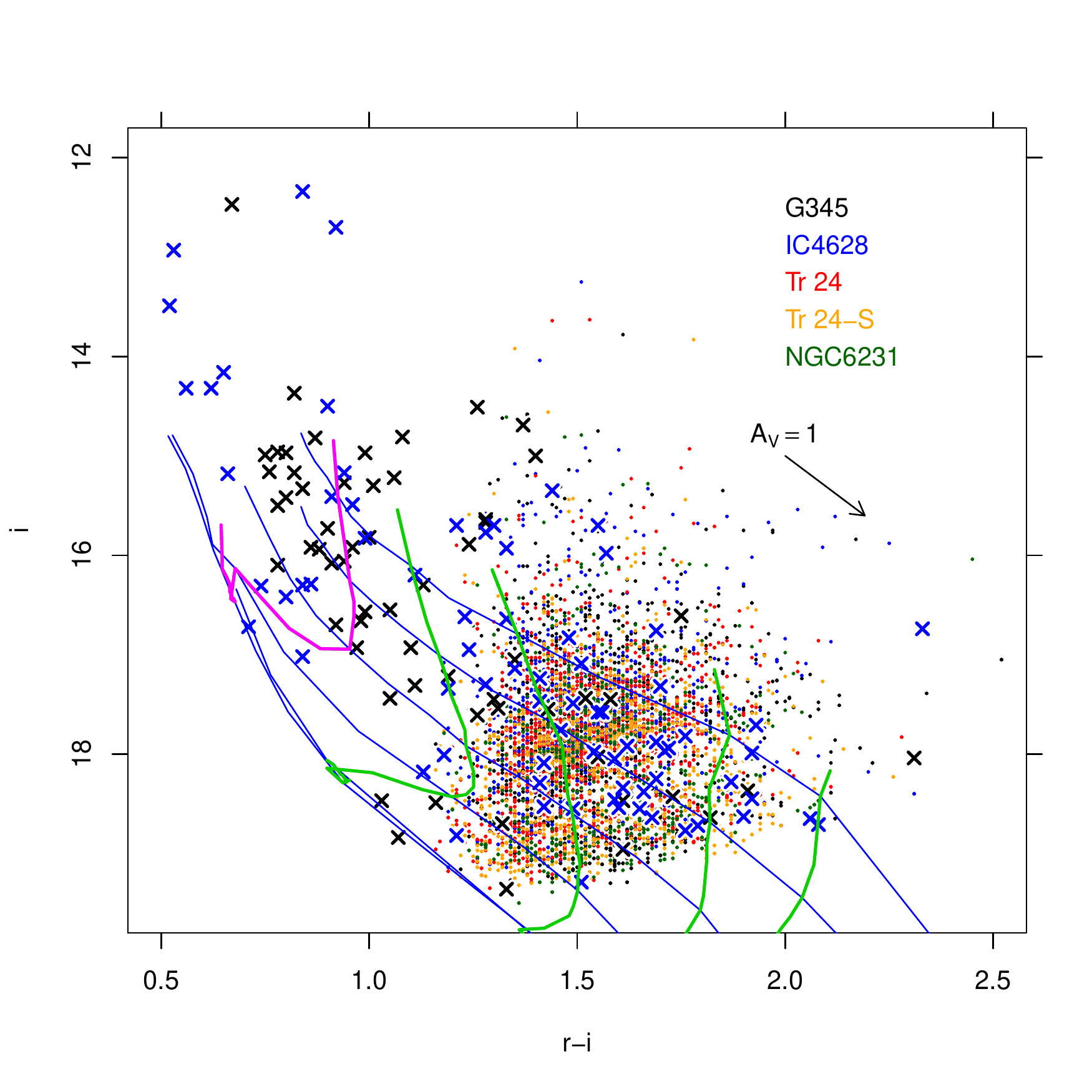}}
\caption{
The $(i,r-i)$ CMD of all M stars, color-coded by spatial region.
Also shown are X-ray detections from XMM-Newton (black crosses) and
Chandra (blue crosses).
Tracks and isochrones as in Figure~\ref{cmd}.
\label{cmd-mstars}}
\end{figure}

We have also tried to derive ages for the M stars directly from the
optical CMD and BHAC model isochrones (Figure~\ref{cmd-mstars}).
The M stars of all subregions are spread over a wide range of ages,
which dominates over mean age differences between individual subregions.
We show also the X-ray sources detected with Chandra (in the \gal\
region) and XMM-Newton (in Tr~24), which also do not follow narrow
sequences, but are consistent with an age range between 1-10~Myr.
A similarly large age spread was already found from the deep Chandra data on
NGC~6231 (Damiani \e 2016).
As expected, the X-ray data select stars over a wide range of masses,
often above $1 M_{\odot}$.

\section{Discussion}
\label{discuss}

From all above results it seems very likely that NGC~6231, Tr~24, and
the nebulous regions IC4628 and \gal\ are physically related, and belong
all to the same Sco~OB1 SFR (or better, Sco~OB1 star-formation complex),
with each subregion corresponding to an episode of stellar formation
along an ordered South-North sequence, spanning slightly more than 50~pc
in total length.

We have recently determined the total mass of NGC~6231, based on
{a large}
catalog of more than 1600 members, as $4380 M_{\odot}$
(Damiani \e 2016)\footnote{
Kuhn \e (2017a), using a slightly deeper catalog, derive a similar
cluster mass of $3300-4200 M_{\odot}$.}.
From Table~4
we derive that the total number of OB stars
in Sco~OB1 subregions {1-4} (88 stars) is 94\% of the OB stars in
NGC~6231. If total mass scales with the number of OB stars, we predict
therefore a mass of $4120 M_{\odot}$ for the \gal, IC4628 and Tr~24
regions collectively, and a total stellar mass of $8500 M_{\odot}$ for
the entire Sco~OB1 complex.
If there is a real excess of M stars in Tr~24, as discussed above, the
estimated mass would be significantly larger, since M stars in the mass
range $0.2-0.5 M_{\odot}$ constitute $\sim 20$\% of the total mass,
assuming the IMF from Weidner \e (2010).
{Still, the above mass estimate of $8500 M_{\odot}$}
would make the Sco~OB1 complex the tenth most
massive SFR in the Galaxy, according to the Weidner \e (2010) list.

NGC~6231 is both the oldest and most massive subregion in Sco~OB1.
Assuming an age of
5-7~Myr for NGC~6231 (Sung \e 2013, Damiani \e 2016;
{Kuhn \e 2017a find a slightly younger median age of $\sim 3.3$~Myr})
and 1~Myr for \gal,
the observed age sequence in Sco~OB1 would suggest a star-formation
episode every $\sim 1$~Myr, spaced by $\sim 10$~pc. If
they were caused by some form of triggering, the perturbation
was moving at about 10~km/s, close to the sound speed inside
an \hii\ region. In earlier times, NGC~6231 undoubtedly
created a bright \hii\ region, whose remains are
still observable as the Gum~55 nebula, a weak \ha\ ring
($5^{\circ}$ in diameter) centered on NGC~6231 (Reipurth 2008).
Why this wave of star formation proceeded only in one direction, and why
it caused distinct episodes instead of a continuous stream of new stars,
remains completely unclear.
{Also needing better clarification are the different
age ranges found by Sung \e (2013) between high- and low-mass stars in NGC~6231
(4-7~Myr and 1-7~Myr, respectively).}

The different morphologies of the spatial distribution of PMS stars in
the individual subregions are also remarkable. NGC~6231 has the most
regular shape, with a (nearly) circular symmetry (Damiani \e 2016;
{more detail about the cluster morphology is available from Kuhn
\e 2017b}). At
the opposite extreme, PMS stars in \gal\ are regrouped in small
subclusters: two are seen in Figure~\ref{acis}, and a few more in
Figure~\ref{spatial-Mstars} (lower-right panel). Also in the IC4628
region PMS stars (especially those with \ha\ and NIR excess) tend to
form small clumps. Tr~24 has a very wide spatial distribution, which was
already found as remarkable by Seggewiss (1968), and somewhat anomalous
by Perry \e (1991). These latter authors argued that the existence of
the Tr~24 cluster itself was questionable, on the basis of its
atipicity, while our results indicate clearly that it exists, and has a
large spatial size.

While dynamical processes have surely erased any signatures of its
original spatial distribution, we may by analogy assume that at birth
Tr~24 was a discrete collection of clumps of stars, like in the \gal\
and IC4628 regions (or, for example, like the subgroups of PMS stars in the
NGC~7000/IC~5070 complex, see Damiani \e 2017). This would explain the
lack of a definite center among its OB stars, which were regrouped by
Seggewiss (1968) in three different subclusters.
The more mobile M stars do instead define a clear center
(Figure~\ref{smoothed-Mstars}).
Alternative explanations such as a rapid expansion after gas removal
(in a shorter time than NGC~6231) seem less likely, since the relatively
few OB stars in Tr~24 would have taken more time to disperse the
surrounding cloud than the time taken by the NGC~6231 OB stars to
disperse their own.
Since little dust is found in the Tr~24 direction, and Tr~24 OB stars
are comparatively few, this cluster probably formed in a lower-density
environment compared to NGC~6231.

Containing less mass in a wider spatial region, Tr~24 is probably less
bound than NGC~6231 (a confirmation would need better kinematical data
than available today). Therefore, it might be a good candidate as a
cluster in the early dissolution phase (Lada and Lada 2003): as discussed in
Section~\ref{intro}, most clusters are predicted not to survive after 10~Myr,
but no clear observational evidence exists of the dissolution process
itself. As PMS stars become more and more spread over large spatial regions,
the chances of identifying them as a population decrease rapidly
with decreasing density contrast with the field-star population.
Our method for finding large numbers of low-mass PMS star candidates over
large sky areas is therefore very promising for an observational study
of this transitional phase of cluster evolution.

It is often stated, and confirmed by kinematical studies, that stars in
a cluster share a common space motion, inherited from the parent
molecular cloud. In the case of two clusters, such as NGC~6231 and
Tr~24, which appear to have originated from the same cloud (they
are also considered as a double cluster by Korkov and Orlov 2011) only a
few Myr one after another, this may lead to a rapid collapse if the cloud
was not rotating, since the two clusters will fall toward their common
barycenter. Assuming no residual gas, and two clusters with the
above-inferred masses and distances for NGC~6231 and Tr~24,
respectively, collision would occur after $\sim 17$~Myr, of which the age
of these clusters is an appreciable fraction. Since Tr~24 is so extended
and NGC~6231 much more massive and compact, we would expect tidal tails
to develop on both sides of Tr~24, toward and away from NGC~6231.
Stars in the Tr~24-S region might in fact constitute one of these tails.
Their distribution in Figure~\ref{smoothed-Mstars} does not point
directly toward NGC~6231, but is suggestive of rotation with respect to
the inter-cluster axis. A rotating parent molecular cloud is more
likely than a non-rotating one, and this might have prevented free fall
of one cluster toward the other (which is also ruled out by the results
from Laval 1972b, indicating expansion). A rotating cloud might also explain
the curved shape of the whole Sco~OB1 complex.
Moreover, Laval 1972a finds that gas in the North and South of
Sco~OB1 moves at different radial velocities.
Applying Kepler's third law to two idealized point masses of
order of NGC~6231 and the rest of Sco~OB1, with a semi-axis of 12~pc,
and a barycenter not far from NGC~6231 itself, which contains nearly
half the total mass of the complex,
gives a rough estimate of a rotation period of 40~Myr. During the $\sim
6$~Myr since their formation, the system would have rotated by $\sim
54^{\circ}$, which is close to the bending angle of the star
distribution seen in Figure~\ref{smoothed-Mstars}.

Finally, we consider the youngest region comprising \gal\ and IC4628,
which has a very different appearance in the mid-IR (Figure~\ref{wise})
and in the optical (Figure~\ref{dss-red}). A better figure of the IC4628
nebula is actually shown by Reipurth (2008; Fig.8): the ionization pattern 
indicates clearly that its ionizing source lies South (or southwest) of
the nebula, in agreement with Crampton and Thackeray (1971).
Ionization in turn causes increased gas compression according to the
Stromgren theory.
On the other hand, the WISE image of Figure~\ref{wise} shows an arc-like
region, suggestive of being pushed by a source in the North, close to
IRAS~16562-3959. Since the
two images, according to our unified view, show just two components of
the same physical cloud, namely heated dust and ionized gas, the cloud
medium is being compressed from both sides, which may be expected to
lead to a rapid density increase, and enhanced star formation
(see Gaczkowski \e 2015) as indicated by our PMS candidates.

\section{Conclusions}
\label{concl}

This study has revealed a large population of candidate PMS stellar
members of the Sco~OB1 association, on the basis of data from the
wide-area VPHAS+ and VVV photometric surveys. In particular, the
combination of a large studied region, deep and uniform multi-band
photometry, and a carefully devised color selection has permitted to
uncover about 4000 M-type candidate Sco~OB1 members,
spread over the entire association.
Although not complete in mass {(both towards solar-like masses, and
towards the lowest stellar masses)}, this is the largest sample of
candidate members of this star-formation region ever selected.

The individual clusters NGC~6231 and Tr~24, and the young \hii\ regions
IC4628 and \gal\ have been shown to be subregions of the same
star-forming Sco~OB1 complex, falling along a {likely} ordered age sequence
spanning almost 10~Myr, from the oldest NGC~6231 cluster to the youngest \gal\
cloud. The entire size of the complex is slightly over 50~pc. The
distribution of member stars has an arc-like shape, which is consistently
recovered using any of the available tracers (M stars, NIR-excess,
UV-excess, \ha-emission stars, OB stars, X-ray sources).
Five major stellar aggregates are found, coincident with the \hii\ regions
\gal\ and IC4628, and clusters Tr~24, its southern extension, and NGC~6231,
respectively, from North to South.
The inferred total mass of $\sim 8500 M_{\odot}$ places
the Sco~OB1 complex among the most massive star-forming complexes in
the Galaxy.

There is a remarkable difference between the compactness of the NGC~6231
cluster and the sparseness of its neighbor Tr~24. This latter was
probably formed in a lower-density environment than NGC~6231, on the
basis of the relative amount of residual dust and its dispersal
timescales. Moreover, Tr~24 may be a good candidate as a cluster in the
early dissolution phase, which our analysis method is particularly
suitable to detect. The presence of the dense and massive cluster NGC~6231
nearby may be accelerating the Tr~24 dissolution process via tidal
effects, of which we also find traces. In the northern parts of the
Sco~OB1 complex, star formation occurs predominantly in smaller-scale
groups, found from both optical and NIR PMS candidates and X-ray data.

The non-member contamination level derived for our M-type PMS candidates
is found in the range 37-51\%, averaged over wide regions.
Locally, it depends on the richness and density of subclusters, and reddening.
The average contamination is larger than than estimated from X-ray
studies (5-10\%), but the comparison is vitiated by the fact that these
latter only refer to the denser cluster centers.
X-ray studies with the sensitivity needed to detect M stars across the
entire Sco~OB1 complex
would require an enormous observing time with current X-ray telescopes,
and are therefore unfeasible in practice.
Therefore, our technique is a valid substitute for the study of PMS
populations covering wide
sky regions, and is sensitive enough to be applied to star-forming
regions as distant as the Sagittarius arm, like Sco~OB1.
It is thus also a promising tool for PMS cluster studies using
next-generation photometric surveys (e.g., LSST).

We also suggest that the curved shape of the Sco~OB1 complex may be
related to its global rotation, inherited from the parent cloud.
The upcoming data from the Gaia satellite will permit to test this
hypothesis.

\begin{appendix}

\section{VVV-2MASS photometric calibration}
\label{append1}

Using the set of sources common to both 2MASS and VVV catalogs, we compared
their respective magnitudes and colors, to put all sources from our final,
combined catalog in the same system. There are slight but non-negligible
differences in the photometry from 2MASS and VVV, which can be approximated
satisfactorily using linear models. We consider only magnitudes with errors
less than 0.07~mag, and colors with errors less than 0.1~mag.

\begin{figure}
\resizebox{\hsize}{!}{
\includegraphics[angle=90]{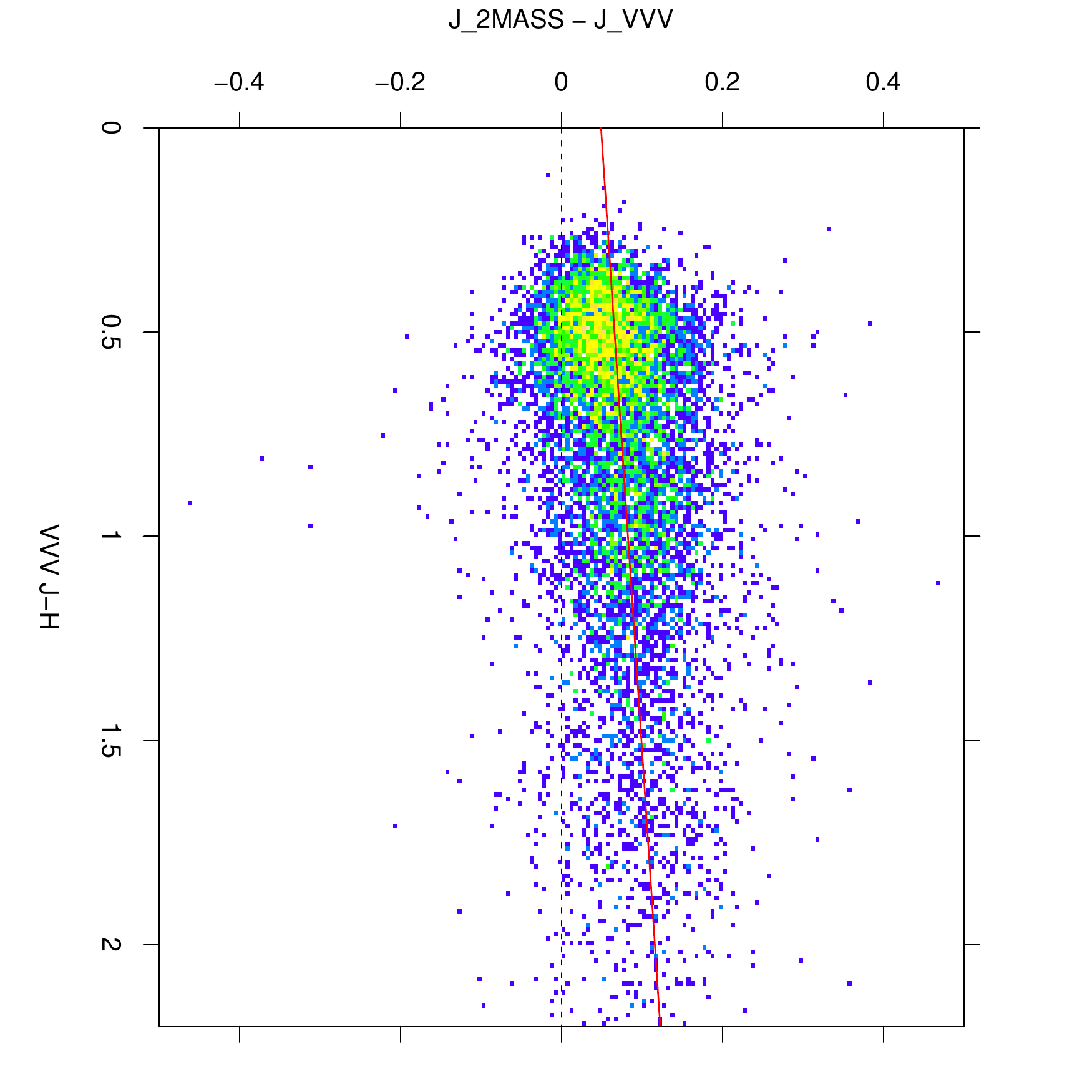}}
\caption{
2MASS-VVV comparison: difference $J_{2MASS} - J_{VVV}$ vs.\
VVV $J-H$ color (2-d histogram).
The black dashed line represents $J_{2MASS} - J_{VVV}=0$, while the
red solid line is a best-fit to the data.
\label{calibr-jj-jh}}
\end{figure}

The comparison between 2MASS and VVV $J$ magnitudes, as a function of
VVV $J-H$ color, is shown in Figure~\ref{calibr-jj-jh}.
A least-squares fit to the data has the form:
\begin{equation}
J_{2MASS} - J_{VVV} = 0.04902  +0.03344 \; (J-H)_{VVV}.
\end{equation}

\begin{figure}
\resizebox{\hsize}{!}{
\includegraphics[angle=90]{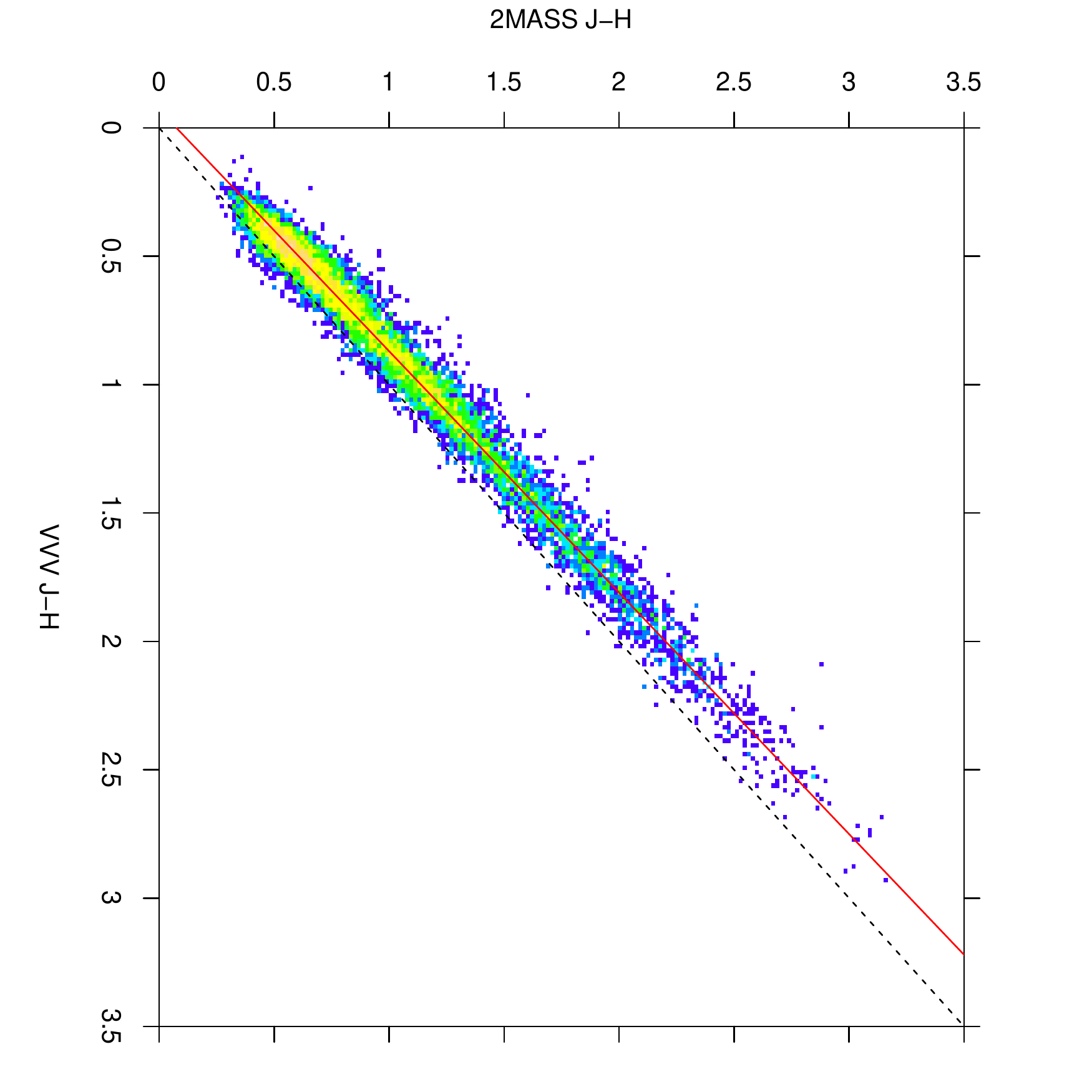}}
\caption{
Comparison between $J-H$ colors from 2MASS and VVV.
The black dashed line represents $(J-H)_{2MASS}=(J-H)_{VVV}$, while
the red solid line is a best-fit to the data.
\label{calibr-jh-jh}}
\end{figure}

\begin{figure}
\resizebox{\hsize}{!}{
\includegraphics[angle=90]{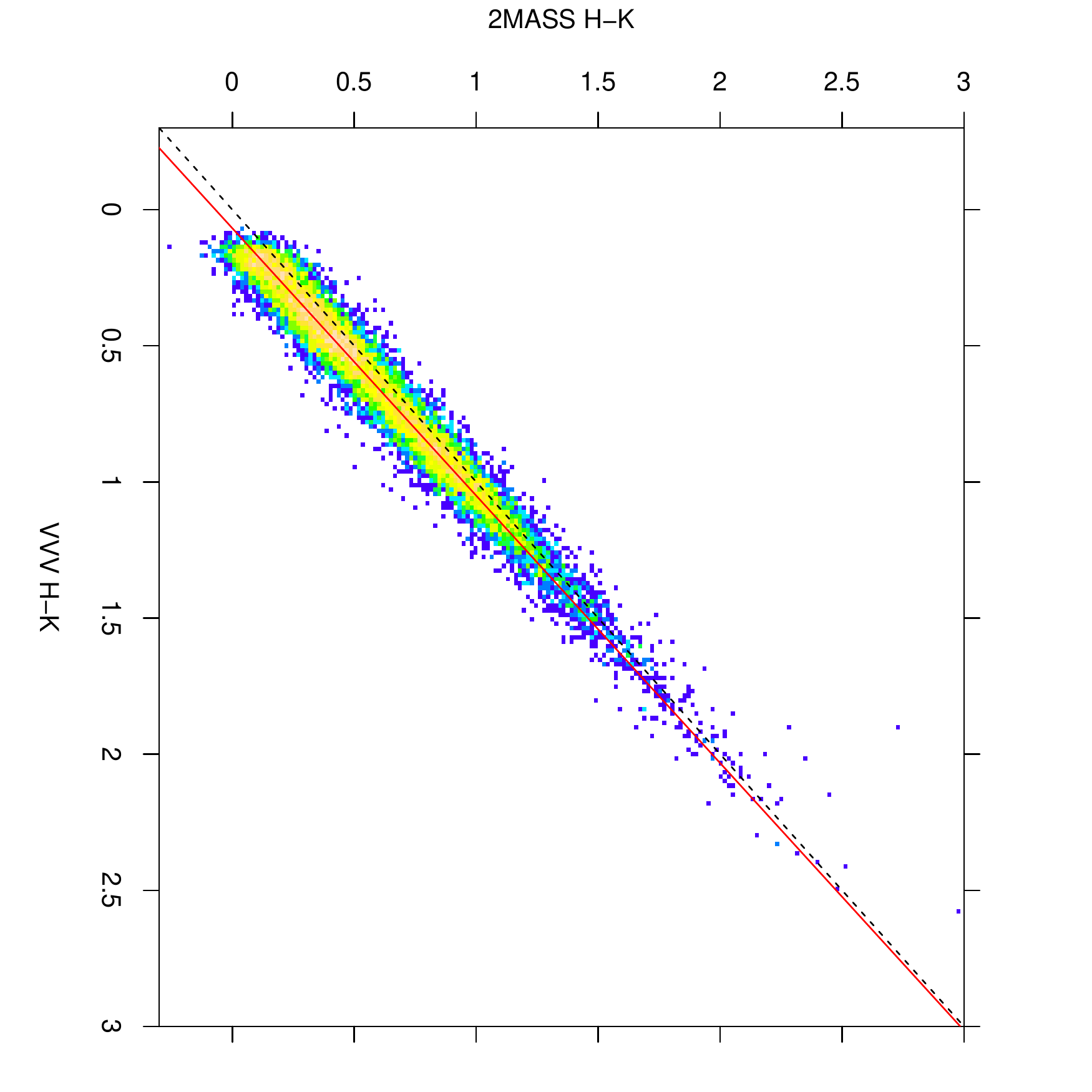}}
\caption{
Same as Fig.~\ref{calibr-jh-jh}, for $H-K$.
\label{calibr-hk-hk}}
\end{figure}

The 2MASS and VVV $J-H$ colors are directly compared in
Figure~\ref{calibr-jh-jh}.
The linear best fit to the data has the form:
\begin{equation}
(J-H)_{2MASS} = 0.07523 +1.06333 \; (J-H)_{VVV}.
\end{equation}
Similarly, Figure~\ref{calibr-hk-hk} shows the comparison between the
respective $H-K$ colors, with the best fit here given by:
\begin{equation}
(H-K)_{2MASS} = -0.06954  +1.01811 \; (H-K)_{VVV}.
\end{equation}

Using these formulae, $JHK$ magnitudes from the VVV catalog were converted to
the 2MASS photometric system.

\end{appendix}

\begin{acknowledgements}
We wish to thank an anonymous referee for his/her helpful suggestions.
Based on data products from observations made with ESO Telescopes at the
La Silla Paranal Observatory under programme ID 177.D-3023, as part of
the VST Photometric H$\alpha$ Survey of the Southern Galactic Plane and Bulge
(VPHAS+, www.vphas.eu).
Also based on data products from VVV Survey observations made with the VISTA
telescope at the ESO Paranal Observatory under programme ID 179.B-2002.
This publication makes use of data products from the Wide-field Infrared
Survey Explorer, which is a joint project of the University of
California, Los Angeles, and the Jet Propulsion Laboratory/California
Institute of Technology, funded by the National Aeronautics and Space
Administration.
The scientific results reported in this article are also based
on observations made by the Chandra and XMM-Newton X-ray Observatories.
This research makes use of the SIMBAD database and the Vizier catalog service,
operated at CDS, Strasbourg, France.
We also make heavy use of R: A language and environment for statistical
computing. R Foundation for Statistical Computing, Vienna, Austria.
(http://www.R-project.org/).
\end{acknowledgements}

\bibliographystyle{aa}

\begin{landscape}
\begin{table}
\centering
\caption{Chandra ACIS-I X-ray sources in \gal. Full table in electronic
format only.
\label{table-acis}} 
\begin{tabular}{cccccccccccccc}
  \hline
 X-ray & CXO Id & RA & Dec & Pos.\ err.\  & Count rate & Rate error &
 VPHAS Id & $r$ & $g-r$ & $r-i$ & $J$ & $J-H$ & $H-K$ \\
 no.\ & (CXOU) & (J2000) & (J2000) & (arcsec) & (cts/ksec) & (cts/ksec) &
 & & & & & & \\
  \hline
  1 & J165852.8-400220 & 254.72007 & -40.03892 & 2.66 & 0.724 & 0.143 &  &  &  &  &  &  &  \\ 
    2 & J165852.8-400135 & 254.72033 & -40.02660 & 4.46 & 0.512 & 0.172 &  &  &  &  & 15.74 & 1.49 & 0.65 \\ 
    3 & J165855.1-400224 & 254.72991 & -40.04025 & 1.58 & 0.395 & 0.147 & J165855.1-400224.2 &  &  &  &  &  &  \\ 
    4 & J165856.4-400530 & 254.73521 & -40.09175 & 1.87 & 1.869 & 0.195 & J165856.4-400530.4 & 15.40 & 1.59 & 0.90 & 12.95 & 0.76 & 0.22 \\ 
    5 & J165857.5-400507 & 254.73975 & -40.08551 & 1.33 & 0.685 & 0.121 &  &  &  &  &  &  &  \\ 
    6 & J165858.8-400150 & 254.74519 & -40.03072 & 1.91 & 0.229 & 0.101 & J165858.7-400150.2 &  &  &  & 10.08 & 0.62 & 0.22 \\ 
    7 & J165900.9-400149 & 254.75401 & -40.03036 & 1.98 & 0.544 & 0.159 & J165901.0-400149.0 & 16.82 &  & 0.99 &  &  &  \\ 
    8 & J165901.9-400219 & 254.75827 & -40.03868 & 1.51 & 0.376 & 0.122 & J165901.9-400219.4 & 20.04 &  & 1.66 &  &  &  \\ 
    9 & J165903.3-400535 & 254.76408 & -40.09307 & 3.10 & 0.344 & 0.116 & J165903.1-400529.6 & 20.03 &  & 1.21 & 17.05 & 0.86 & 0.29 \\ 
   10 & J165903.7-400635 & 254.76555 & -40.10978 & 1.80 & 0.189 & 0.075 &  &  &  &  &  &  &  \\ 
   11 & J165904.5-400617 & 254.76886 & -40.10489 & 1.40 & 0.505 & 0.146 & J165904.4-400618.1 & 18.53 & 2.27 & 1.19 & 15.06 & 0.95 &  \\ 
   12 & J165905.2-400614 & 254.77194 & -40.10389 & 1.33 & 0.150 & 0.074 &  &  &  &  &  &  &  \\ 
   13 & J165905.6-400623 & 254.77345 & -40.10654 & 1.01 & 1.252 & 0.151 & J165905.6-400623.6 &  & 0.68 &  & 11.78 & 0.28 & 0.31 \\ 
   14 & J165905.8-395927 & 254.77433 & -39.99090 & 1.94 & 0.179 & 0.083 & J165905.9-395924.9 &  &  &  &  &  &  \\ 
   15 & J165906.0-400708 & 254.77540 & -40.11916 & 1.58 & 0.324 & 0.107 &  &  &  &  & 14.96 & 1.00 & 0.48 \\ 
   16 & J165906.2-400600 & 254.77625 & -40.10003 & 1.30 & 0.533 & 0.106 &  &  &  &  &  &  &  \\ 
   17 & J165906.8-395903 & 254.77836 & -39.98424 & 1.94 & 0.583 & 0.166 &  &  &  &  &  &  & 0.23 \\ 
   18 & J165907.4-400912 & 254.78116 & -40.15356 & 1.84 & 0.271 & 0.134 & J165907.4-400913.3 & 18.88 &  & 1.41 &  &  & 0.30 \\ 
   19 & J165908.0-400537 & 254.78334 & -40.09385 & 1.01 & 0.511 & 0.151 &  &  &  &  & 16.75 & 2.78 & 1.62 \\ 
   20 & J165908.6-401037 & 254.78591 & -40.17699 & 1.98 & 0.311 & 0.170 & J165908.5-401038.3 & 20.15 &  & 1.87 & 15.07 & 1.30 & 0.51 \\ 
   \hline
\end{tabular}
\end{table}
\end{landscape}

\begin{landscape}
\begin{table}
\centering
\caption{XMM-Newton EPIC X-ray sources in Tr~24. Full table in electronic
format only.
\label{table-xmm}} 
\begin{tabular}{cccccccccccccc}
  \hline
X-ray & XMM Id & RA & Dec & Pos.\ err.\  & Count rate & Rate error &
VPHAS Id & $r$ & $g-r$ & $r-i$ & $J$ & $J-H$ & $H-K$ \\
no.\ & (XMMU) & (J2000) & (J2000) & (arcsec) & (cts/ksec) & (cts/ksec) &
& & & & & & \\
  \hline
385 & J165422.2-412148 & 253.59279 & -41.36351 & 3.89 & 15.777 & 3.439 &  &  &  &  &  &  &  \\ 
  386 & J165442.1-412056 & 253.67548 & -41.34911 & 5.76 & 5.784 & 1.363 &  &  &  &  & 9.21 & 0.04 & 0.05 \\ 
  387 & J165500.0-412041 & 253.75034 & -41.34476 & 2.77 & 3.579 & 0.781 & J165500.2-412039.6 & 17.23 & 2.14 &  &  &  &  \\ 
  388 & J165428.6-412040 & 253.61951 & -41.34454 & 5.44 & 7.007 & 1.474 & J165428.8-412040.4 & 13.54 & 1.07 &  & 11.82 & 0.59 & 0.55 \\ 
  389 & J165502.8-412009 & 253.76187 & -41.33610 & 5.36 & 5.795 & 1.192 & J165502.9-412010.7 &  &  &  & 14.63 & 0.71 & 0.17 \\ 
  390 & J165526.2-412002 & 253.85930 & -41.33398 & 4.03 & 3.451 & 0.800 & J165526.3-412001.0 &  &  &  &  &  &  \\ 
  391 & J165427.6-411941 & 253.61530 & -41.32824 & 1.44 & 3.014 & 0.718 &  &  &  &  &  &  &  \\ 
  392 & J165419.6-411937 & 253.58170 & -41.32699 & 4.28 & 9.853 & 1.326 & J165419.7-411935.7 &  &  &  &  &  &  \\ 
  393 & J165453.7-411937 & 253.72409 & -41.32698 & 5.94 & 3.721 & 0.925 & J165453.6-411936.1 & 17.00 & 1.72 & 0.94 & 14.60 & 0.78 & 0.21 \\ 
  394 & J165524.5-411830 & 253.85210 & -41.30859 & 4.43 & 2.464 & 0.676 & J165524.7-411835.2 & 19.03 &  & 1.58 &  &  &  \\ 
  395 & J165415.4-411754 & 253.56430 & -41.29852 & 4.03 & 2.976 & 0.688 & J165415.5-411755.2 & 20.46 &  & 1.82 &  &  &  \\ 
  396 & J165502.6-411754 & 253.76104 & -41.29834 & 4.75 & 5.857 & 0.960 &  &  &  &  &  &  &  \\ 
  397 & J165420.7-411743 & 253.58646 & -41.29539 & 2.88 & 8.424 & 1.023 &  &  &  &  &  &  &  \\ 
  398 & J165410.8-411724 & 253.54540 & -41.29024 & 2.70 & 3.414 & 0.717 & J165410.8-411726.3 & 19.50 & 1.96 & 1.03 &  &  &  \\ 
  399 & J165502.6-411714 & 253.76113 & -41.28731 & 2.84 & 2.528 & 0.572 & J165502.7-411714.1 & 17.30 & 1.97 &  & 14.42 & 0.76 & 0.28 \\ 
  400 & J165440.0-411645 & 253.66679 & -41.27940 & 4.72 & 5.322 & 0.866 &  &  &  &  & 18.73 & 1.24 & 0.38 \\ 
  401 & J165443.0-411629 & 253.67922 & -41.27477 & 2.95 & 7.478 & 0.954 &  &  &  &  & 18.09 & 0.93 & 0.48 \\ 
  402 & J165354.9-411606 & 253.47888 & -41.26858 & 3.38 & 85.949 & 7.321 &  &  &  &  &  &  &  \\ 
  403 & J165409.4-411546 & 253.53954 & -41.26296 & 2.45 & 4.322 & 0.758 & J165409.5-411545.0 & 13.04 & 0.91 &  & 11.76 & 0.32 & 0.12 \\ 
  404 & J165533.4-411540 & 253.88923 & -41.26133 & 2.27 & 1.454 & 0.420 & J165533.5-411539.8 & 16.99 & 1.66 & 0.91 &  &  &  \\ 
   \hline
\end{tabular}
\end{table}
\end{landscape}

\begin{landscape}
\begin{table}
\centering
\caption{Optical and NIR photometry for M stars, and stars with UV or IR
excess, or H$\alpha$ emission. Full table in electronic format only.
\label{table-mstars}}
\begin{tabular}{rcccccccccccccc}
  \hline
Seq & RA & Dec & VPHAS Id & $i$ & $r-i$ & $r-H\alpha$ & $g-r$ & $u-g$ &
$J$ & $J-H$ & $H-K$ & IR & \ha\ & UV \\
no.\ & (J2000) & (J2000) & & & & & & & & & & excess & emission & excess \\
  \hline
1 & 253.10967 & -42.49575 & J165226.3-422944.6 & 19.61 & 1.37 & 0.74 &  &  & 16.55 & 0.75 & 0.16 &  &  &  \\ 
  2 & 253.22413 & -42.47575 & J165253.8-422832.6 & 17.33 & 0.79 & 0.37 &  &  & 15.11 & 0.82 & 0.44 & Y &  &  \\ 
  3 & 253.21472 & -42.45311 & J165251.5-422711.2 & 17.18 & 0.91 & 0.44 &  &  & 14.44 & 1.47 & 0.63 & Y &  &  \\ 
  4 & 253.43352 & -42.43837 & J165344.0-422618.1 & 19.54 & 1.51 & 0.78 &  &  & 16.07 & 0.82 &  &  &  &  \\ 
  5 & 253.74475 & -42.46394 & J165458.7-422750.3 & 19.00 & 1.11 & 0.35 &  &  & 14.86 & 1.49 & 0.66 & Y &  &  \\ 
  6 & 253.99296 & -42.46397 & J165558.3-422750.3 & 18.17 & 1.00 & 0.38 &  &  & 15.34 & 1.02 & 0.63 & Y &  &  \\ 
  7 & 254.03561 & -42.45187 & J165608.5-422706.7 & 19.09 & 1.34 & 0.64 &  &  & 16.13 & 0.78 & 0.15 &  &  &  \\ 
  8 & 254.15191 & -42.47327 & J165636.5-422823.8 & 18.51 & 1.03 & 0.42 &  &  & 15.45 & 1.19 & 0.77 & Y &  &  \\ 
  9 & 254.28706 & -42.48321 & J165708.9-422859.7 & 20.19 & 1.13 &  &  &  & 16.85 & 1.94 & 1.18 & Y &  &  \\ 
  10 & 254.36392 & -42.46472 & J165727.4-422753.0 & 19.44 & 1.09 & 0.35 &  &  & 16.07 & 1.40 & 0.75 & Y &  &  \\ 
  11 & 254.39702 & -42.48811 & J165735.3-422917.1 & 18.74 & 0.98 & 0.37 &  &  & 16.07 & 0.73 & 0.51 & Y &  &  \\ 
  12 & 254.65041 & -42.49495 & J165836.1-422941.7 & 19.28 & 1.25 & 0.35 &  &  & 15.59 & 1.04 & 0.74 & Y &  &  \\ 
  13 & 254.81260 & -42.45834 & J165915.0-422730.0 & 18.26 & 1.25 & 0.47 &  &  &  &  & 0.60 & Y &  &  \\ 
  14 & 254.94841 & -42.49392 & J165947.6-422938.1 & 19.77 & 1.42 & 0.50 &  &  & 14.63 & 2.46 & 1.13 & Y &  &  \\ 
  15 & 254.98341 & -42.44989 & J165956.0-422659.5 & 19.49 & 1.22 & 0.45 &  &  & 15.38 & 1.92 & 1.06 & Y &  &  \\ 
  16 & 252.69726 & -42.41120 & J165047.3-422440.3 & 17.94 & 1.28 &  &  &  & 15.03 & 0.86 & 0.25 &  &  &  \\ 
  17 & 253.18205 & -42.39847 & J165243.7-422354.5 & 20.27 & 1.75 &  &  &  & 16.57 & 0.63 & 0.20 &  &  &  \\ 
  18 & 253.19191 & -42.40229 & J165246.1-422408.3 & 15.78 & 0.86 & 0.33 &  &  & 13.46 & 0.64 & 0.49 & Y &  &  \\ 
  19 & 253.32115 & -42.37896 & J165317.1-422244.2 & 20.08 & 1.61 &  &  &  & 16.48 & 0.80 & 0.32 &  &  &  \\ 
  20 & 253.40177 & -42.38816 & J165336.4-422317.4 & 19.82 & 1.19 & 0.33 &  &  & 16.30 & 1.03 & 0.69 & Y &  &  \\ 
   \hline
\end{tabular}
\end{table}
\end{landscape}

\end{document}